# FROM Si NANOWIRES TO Ge NANOCRYSTALS FOR VIS-NIR-SWIR SENSORS AND NON-VOLATILE MEMORIES: A REVIEW

Ana-Maria LEPADATU[1,*], Ionel STAVARACHE[1], Catalin PALADE[1], Adrian SLAV[1], Valentin A. MARALOIU[1], Ioana DASCALESCU[1], Ovidiu COJOCARU[1], Valentin S. TEODORESCU[1,2], Toma STOICA[1], Magdalena L. CIUREA[1,2,*]

**Rezumat.** *Si și Ge nanocristalin prezintă un interes crescut pentru fotonica integrată pe Si cu aplicații în dispozitive de emisie de lumină, senzori optici, fotodetectori, captarea și conversia energiei solare, precum și pentru memorii nevolatile cu poartă flotantă (NVMs). În această lucrare prezentăm filme de Si poros nanocristalin (nc-PS), filme formate din doturi cuantice (QDs) de Si înglobate în matrice de $SiO_2$ și nanocristale (NCs)/QDs/nanoparticule (NPs) de Ge imersate în oxizi ($SiO_2$, $TiO_2$, $HfO_2$, $Al_2O_3$). Cea mai atractivă și importantă proprietate a nc-PS este fotoluminescența intensă în domeniul vizibil (VIS) la temperatura camerei (RT), pe când NCs/NPs de Ge imersate în oxizi prezintă fotosensibilitate crescută în vizibil-infraroșu apropiat-infraroșu de lungimi de undă scurte (VIS-NIR-SWIR). Astfel, spectrele de fotocurent măsurate la RT se extind în SWIR până la 1325 nm. NVMs cu poartă flotantă formate din NCs/NPs/QDs de Ge prezintă proprietăți de memorie de înaltă performanță, caracteristicile de retenție corespunzând datelor actuale raportate în literatură pentru NVMs cu poartă flotantă pe bază de NCs. De asemenea, demonstrăm importanța controlului parametrilor de preparare pentru obținerea filmelor cu proprietăți țintite de fotoluminescență, fotosensibilitate și proprietăți de stocare de sarcină pentru aplicații, e.g. senzori optici și fotodetectori în VIS-NIR-SWIR, precum și NVMs electronice și fotoelectrice. Evidențiem corelația dintre condițiile de preparare, morfologia, compoziția și structura cristalină cu proprietățile optice, electrice, fotoelectrice și de stocare de sarcină, și de asemenea evidențiem contribuția efectului de confinare cuantică, a stărilor localizate și a centrilor de captură.*

**Abstract.** *Nanocrystalline Si and Ge are of high interest for integrated Si photonics related to light emission, optical sensors, photodetectors, solar energy harvesting and conversion devices, and also for floating gate non-volatile memories (NVMs). In this review, we have focused on nanocrystalline porous Si (nc-PS) with extension to Si nanodots, and Ge nanocrystals (NCs)/quantum dots (QDs)/nanoparticles (NPs) embedded in oxides ($SiO_2$, $TiO_2$, $HfO_2$, $Al_2O_3$). The great asset of nc-PS is its intense photoluminescence in VIS at room temperature (RT), while Ge NCs/NPs embedded in oxides show high photosensitivity in VIS-NIR-SWIR in the spectral photocurrent up to*

---

[1] National Institute of Materials Physics, 405A Atomistilor Street, 077125 Magurele, Romania
[2] Academy of Romanian Scientists, 54 Splaiul Independentei, 050094 Bucuresti, Romania
* Correspondence: ciurea@infim.ro, lepadatu@infim.ro

*1325 nm at RT. Ge NCs/NPs/QDs floating gate NVMs present high memory performance, the retention characteristics corresponding to the state of the art for NCs floating gate NVMs. We prove the relevance of controlling the preparation parameters for obtaining films with targeted photoluminescence, photosensitivity and charge storage properties for applications, e.g. VIS-NIR-SWIR optical sensors and photodetectors, and electronic and photoelectric NVMs. We evidence the correlation of preparation conditions, morphology, composition and crystalline structure with optical, electrical, photoelectrical and charge storage properties and also evidence the contribution of quantum confinement effect, localized states and trapping centers.*

**Keywords:** nanocrystalline porous Si, Si nanodots, Ge nanocrystals embedded in oxides, VIS-NIR-SWIR sensors, floating gate non-volatile memories

## 1. Introduction

Group IV semiconductors are intensively investigated [1-4] along with group III–V [5, 6], II–VI [7-17] and IV–VI semiconductors [18, 19].

From Group IV, Si is the most studied material [20, 21], it is considered by a large research community and engineers in the world as the most "powerful" material for its important applications in micro- and optoelectronics (from discrete devices to integrated circuits), and micromachining. Nanocrystalline Si brings new benefits for these kinds of applications, e.g being nc-PS and films of Si NCs embedded in $SiO_2$ matrix that are an important hope for fabrication of Si based optoelectronic devices. By finding this alternative solution, the III–V and II–VI materials (that are unprotective and unfriendly to the environment) on which at present optoelectronics is based, can be replaced with ecological Si – based materials. nc-PS has many applications in micro- and nanoelectronics, e.g. sacrificial layers, isolation walls etc., as mentioned before.

Special focus is also given to the research of Ge quantum dots (QD)/nanocrystals (NCs) for photodetectors [22-26], solar energy harvesting [27, 28] and conversion devices [29] or for light emission and integrated light sources [30-32], and also for floating gate non-volatile memory devices [33-35]. Ge NCs are compatible with mainstream CMOS technology and present reduced thermal budget compared to Si NCs formation, and also larger exciton Bohr radius than Si.

In this review, we have focused on nanocrystalline porous Si (nc-PS) including Si nanodots, and Ge NCs/QDs/nanoparticles (NPs) embedded in oxides ($SiO_2$, $TiO_2$, $HfO_2$, $Al_2O_3$). The relationship between preparation conditions, morphology, composition and structure, and optical, electrical, photoelectrical and charge storage properties together with the contribution of quantum confinement (QC) effect, trapping levels and localized states are evidenced. We prove the relevance of controlling the preparation parameters for obtaining films with targeted photoluminescence, photosensing and charge storage properties for applications, e.g. visible-near infrared-short wave infrared range (VIS-NIR-SWIR) optical sensors and photodetectors, and floating gate non-volatile memories (NVMs).

## 2. Nanocrystalline Porous Si (nc-PS)

In the early 1990s, Si expanded its area of interest in prestigious research laboratories in the world together with industry by obtaining new nanostructured morphologies as Si nanowires (NWs) and Si nanocrystals (NCs). This significant interest is due to the benefits brought by Si nanostructuring as Si nanowires and nanocrystals have new and tunable properties due to the QC effect [36-43].

Nanocrystalline PS (nc-PS) became of high interest for scientific community 30 years ago and continues to be of interest up to now. It has been intensively investigated by Canham [44] and Lehman and Gösele [45] after evidencing of bright luminescence in visible range (VIS) at room temperature (RT) that is due to radiative recombination between QC levels. This is specific to Si skeleton with nanometric diameters remaining after the anodization process. Depending on skeleton morphology and size, properties of nc-PS can differ very much, and therefore they can be tuned by properly adjusting the preparation parameters, and therefore films with desired properties can be obtained [39, 46-51]. As a consequence, additional applications in optoelectronics for SWIR detection, energy conversion and biomedical applications were evidenced [39, 48, 50, 52, 53].

In this section, we present nc-PS formed of Si NWs after preparation, but these nanowires can be formed by chains of NCs resulted by oxidation that interrupts the Si skeleton.

### 2.1 Preparation of nc-PS films

We obtained nc-PS layers by electrochemical etching (anodization) of p-type (100) Si (5 – 15 Ω·cm resistivity) in HF (49%) – $C_2H_5OH$ solution (1:1 volume ratio) in dark and under constant current density (5 – 25 mA/cm$^2$) conditions [54-58]. After anodization, PS films present a weak photoluminescence (PL), so that a post anodization process was necessary, i.e. a photochemical etching by illuminating them *in situ* (in the same electrolyte) for few minutes (1.5 – 3 min) with a Xenon lamp. Finally, the samples were rinsed in double distilled water and dried in air to obtain fresh samples. These fresh samples were stabilized by both anodic process and storage in air, under normal conditions.

### 2.2 Morphology and structure

The microstructure investigations (transmission electron microscopy (TEM) at low magnification and high resolution TEM (HRTEM)) performed on stabilized films proved that nc-PS films have a double-scale porosity [55, 57]. At the first porosity level, the films show a honeycomb-like system presented in Figure 1a and b, being formed of alveolar columnar pores with 1 – 5 μm diameter separated each to other by walls, this macroporosity being of 70%. The second level of porosity is inside the walls (100 – 200 nm thickness) between alveoli. These walls are formed by a network of NWs with 1 – 5 nm diameter (Figure 1c) and 5 – 35 μm length, having 50% nanoporosity. These NWs form the skeleton of nc-PS (obtained after electrochemical etching) and they keep the crystallinity of bulk Si (0.314 nm lattice fringe) as shown in Figure 1c.

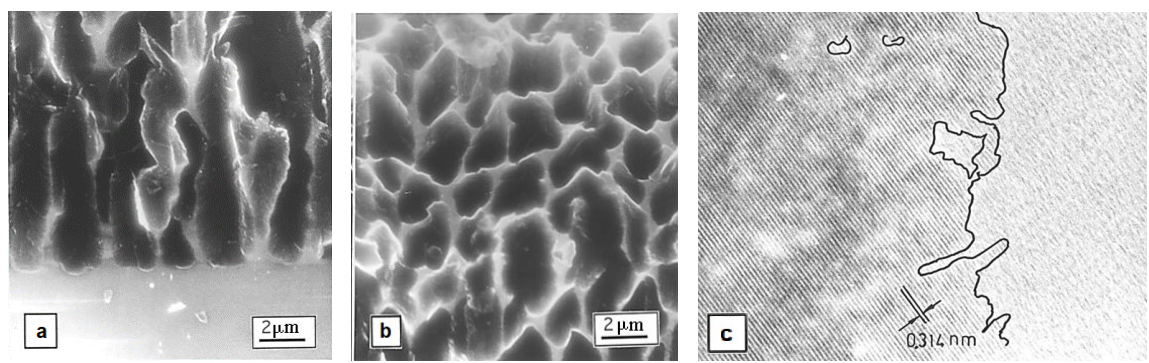

**Fig. 1.** Scanning electron microscopy (SEM, a, b) and HRTEM (c) images: **a.** PS film cross section nearly parallel with the cubic (100) plane of Si; **b.** oblique cross section of PS film, nearly parallel to the (110) cleavage plane of Si; **c.** HRTEM image of the wall surface PS nanostructure revealed by the lattice fringes contrast with respect to the amorphous structure of the silicon oxide and glue [55]. Reproduced with permission from M. L. Ciurea, V. Iancu, V. S. Teodorescu, L. C. Nistor, M. G. Blanchin, Microstructural Aspects Related to Carriers Transport Properties of Nanocrystalline Porous Silicon Films, J. Electrochem. Soc. 146, 3516 (1999), DOI https://doi.org/10.1149/1.1392507 © The Electrochemical Society. Reproduced by permission of IOP Publishing Ltd. All rights reserved.

### 2.3. Electrical transport mechanisms in nc-PS

*Quantum confinement model for energy levels in 1D and 0D systems*

The nanometric PS presents QC effect and therefore a NW (1D system) can be described by a quantum well that generates QC energy levels located in the energy band gap. The QC ground state should coincide at 0 K with the maximum of valence band being the maximum occupied level [57-59]. We consider as a first approximation that Hamiltonian of 1D system can be written as the sum of the longitudinal part – Bloch-like states along the NW, and transversal part described by an infinite rectangular quantum well with cylindrical symmetry. In the effective mass approximation, the electron energy is:

$$E = \varepsilon_{n,k_z} + \frac{2\hbar^2}{m^* d^2} x_{l,p}^2 = \left(\varepsilon_{n,k_z} + \frac{2\hbar^2}{m^* d^2} x_{0,1}^2\right)$$
$$+ \frac{2\hbar^2}{m^* d^2}\left(x_{l,p}^2 - x_{l,p}^2\right) \equiv \varepsilon_{n,k_z}^S + E_{l,p-1} \quad (1)$$

$$E_{l,p} = \frac{2\hbar^2}{m^* d^2}\left(x_{l,p+1}^2 - x_{0,1}^2\right) \quad (2)$$

(reproduced with permission from M. L. Ciurea, *Quantum confinement in nanocrystalline silicon*, J. Optoelectron. Adv. Mater. 7, 2341 (2005)) in which $\varepsilon_{n,k_z}^S$ is the longitudinal Bloch energy and $E_{l,p}$ are the QC levels (transversal part) measured from the maximum of valence band ($E_{0,0} \equiv E_V \equiv 0$), with $m^*$ the transversal effective mass, $m^* = 0.66\, m_e$, $m_e$ – electron mass, $d$ – the average NW diameter and $x_{l,p}$ is the $p$-th zero of the Bessel function $J_l(x)$, $l$ being the orbital quantum number.

This model can be easily applied to a system of Si nanodots ($d \leq 5$ nm) embedded in SiO$_2$ matrix [58, 60-63]. We approximated the shape of Si nanodots with a sphere. In this case, there are only energy quasibands separated by wider forbidden bands that tend to become "direct" because the momentum conservation law is no longer valid (small number of Si atoms in a nanodot). Therefore, the QC energy levels are given by Eq. (2).

As mentioned before, Eqs. (1–2) are valid in the approximation of effective mass ($E \sim d^{-2}$). More rigorous calculations, e.g. linear combination of atomic orbitals, give a more precise dependence of the band gap on diameter $d$. Thus, $E \sim d^{-\alpha}$, $\alpha = 1.02$ for NWs with cylindrical symmetry, $\alpha = 1.39$ for spherical nanodots and $\alpha = 0.6 - 0.8$ resulting from exciton measurements. As the energy differences between QC levels (Eqs. 1–2) are much bigger (at least by one order of magnitude) than the thermal energy $k_BT$, the carrier concentration is given by Boltzmann law, $n \sim \exp(-E_a/k_BT)$, the activation energy $E_a$ being the difference between the last occupied level and the following one. In other words, when a level is filled, the electron transition on the following one will start together with an abrupt change of the activation energy. The ratio of consecutive activation energies is different for QDs than for NWs.

For nanodots, the ratio is:

$$R_d = \frac{E_a''}{E_a'} = \frac{\varepsilon_{l'',p''} - \varepsilon_{l',p'}}{\varepsilon_{l',p'} - \varepsilon_{l,p}} = \frac{x_{l'',p''+1}^2 - x_{l',p'+1}^2}{x_{l',p'+1}^2 - x_{l,p+1}^2} \quad (3)$$

For NWs with cylindrical symmetry, the carriers are always excited from the valence band (electron reservoir), so that:

$$R_w = \frac{x_{l'',p''+1}^2 - x_{0,1}^2}{x_{l',p'+1}^2 - x_{0,1}^2} \tag{4}$$

(reproduced with permission from M. L. Ciurea, *Quantum confinement in nanocrystalline silicon*, J. Optoelectron. Adv. Mater. 7, 2341 (2005))

Considering the excitation conditions, for thermal excitation only (low electric field, $eU \ll k_BT$), the first three levels for cylindrical or spherical symmetries correspond to $l = 0, 1, 2$, while for excitation under a high field ($eU \gg k_BT$) one has angular momentum conservation, the first three levels corresponding to $l = 0$ and $p = 0, 1, 2$.

**In nc-PS films**, the conduction mechanism **in dark** was determined by measuring current-voltage ($I - V$) and current – temperature ($I - T$) characteristics on fresh and stabilized nc-PS films. Fresh nc-PS films present $I - V$ characteristics almost linear at low voltage and slowly rectifying at higher voltage, meaning that electrical transport takes place through Si skeleton, along the NWs [36, 55, 57, 58]. Stabilized nc-PS films present a strong rectifying behavior of $I - V$ curves as Si NWs are (partly) oxidized and they are formed of a chain of Si NCs separated each to other by $SiO_2$, i.e. potential barriers. The $I - T$ characteristics [36] of fresh samples have an Arrhenius behavior with 0.45 – 0.50 eV activation energy on the investigated temperature interval, while $I - T$ curves measured on stabilized films present two activation energies ($E_a$) of $E_{a1} = 0.50 - 0.60$ eV and $E_{a2} = 1.20 - 1.80$ eV at low temperature and high temperature, respectively as shown in Figure 2. The slope change is abrupt and takes place at about 280 K.

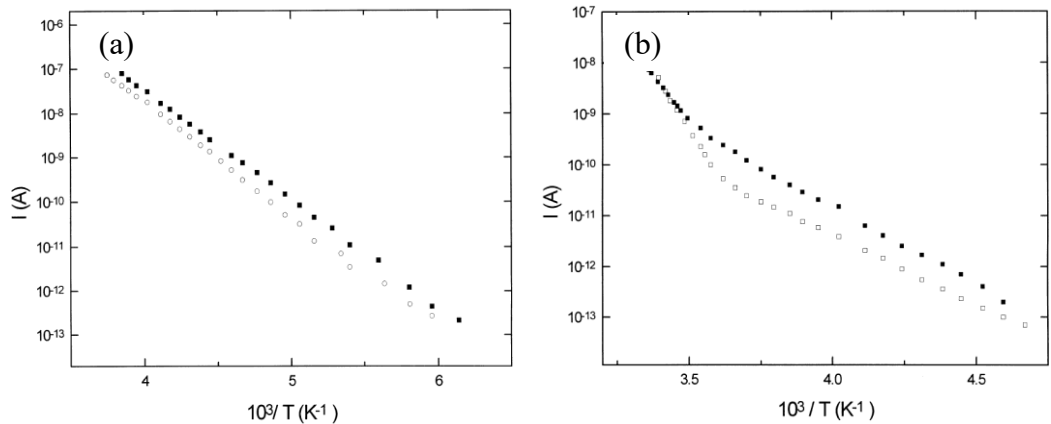

**Fig. 2.** $I - V$ characteristics measured on: **a.** fresh films with 0.45 – 0.50 eV activation energy; **b.** stabilized films with activation energy at low temperature of $E_{a1} = 0.50 - 0.60$ eV and $E_{a2} = 1.20 - 1.80$ eV at high temperature [36]. Reprinted from Solid State Electron. 42, 1893, V. Iancu, M. L. Ciurea, Quantum confinement model for electric transport phenomena in fresh and stored photoluminescent porous silicon films, Copyright © 1998 with permission from Elsevier.

These results were explained in the frame of the QC model, being justified by the Si NW diameters of 3 – 5 nm and lengths of few μm (HRTEM and SEM). By using QC model, the electron transitions responsible for Arrhenius behavior of $I – T$ curves, i.e. for activation energies, can be found. So, the two activation energies obtained in $I – T$ curves corresponding to stabilized films are explained as follows. At a given temperature, the first QC level being filled, the Fermi level will jump on the following one (the second one), so that the activation energy will be changed from $E_{a1}$ to $E_{a2}$. Also, by oxidation Si NW diameters decrease a little, and the band gap will increase correspondingly.

**In the case of films formed of Si NCs embedded in SiO$_2$ matrix** the conduction mechanisms **in dark** were similarly determined as for nc-PS (experimental $I – V$ and $I – T$ curves and applying QC model).

The films morphology of Si NCs in SiO$_2$ with composition of 50% Si: 50% SiO$_2$ is shown in Figure 3. The HRTEM image reveals Si NCs (C1, C2 and C3) and the fringes show their relative crystalline orientation [64].

Like in nc-PS, $I–T$ characteristics taken on this kind of films have an Arrhenius behavior, with two activation energies of 0.30 eV and 0.51 eV at low and higher temperature (for films composition of 73:27).

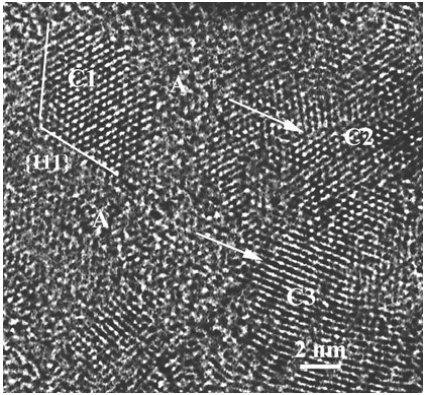

**Fig. 3.** HRTEM image of Si – SiO$_2$ films with 50:50 composition: C1 to C3 are Si NCs and A is amorphous SiO$_2$; relative orientation if Si NC is shown by fringes (C2 and C3 almost parallel, C3 rotated clockwise, ~30º) [64]. Reproduced with permission from I. Stavarache, M. L. Ciurea, Percolation phenomena in Si–SiO$_2$ nanocomposite films, J. Optoelectron. Adv. Mater 9, 2644 (2007).

These characteristics were interpreted in the frame of the QC model (for 0D system of nanodots with 4.5 – 5.5 nm diameter (microstructure investigations). So, the activation energy of 0.30 eV (low temperature) corresponds to a transition between energy levels $E_{0,0}$ and $E_{0,1}$ while the activation energy of 0.51 eV at high temperature corresponds to a transition from $E_{0,1}$ QC level to $E_{0,2}$ one.

The QC model can be also applied in trapping levels investigations, namely to distinguish between trapping levels and QC levels. The trapping levels can also participate in activation energies evidenced in $I - T$ characteristics corresponding to depolarization currents [65-69].

*Visible photoluminescence in nc-PS at room temperature*

The nc-PS has attracted widespread attention of the scientific world due to bright PL in VIS at RT explained by radiative recombination with participation of QC band-edge states [44]. This band is a broad one of 1.40 – 2.20 eV having a long luminescence decay time from µs to ms and it is due to the QC effect in Si nanowires/NCs – if nc-PS films are stabilized by oxidation [39, 60].

We measured the visible PL of PS films obtained by anodization in the dark for different times, followed by chemical etching under illumination (Xe lamp) for different times, too. The PL measurements were performed (excitation with $\lambda$ = 352 nm, $P$ = 1 mW; detector: GaAs photomultiplier: –30 °C, –1500 V; chopping frequency 433 Hz) on films at different stabilization times, i.e. after short term storage (weeks) and after long term storage (months) [54].

The problem we have proposed to solve was to evidence in the PL spectrum, the subbands of the PL broad band. For this we prepared different films anodized and stabilized under different conditions. The corresponding PL spectra are given in Figure 4.

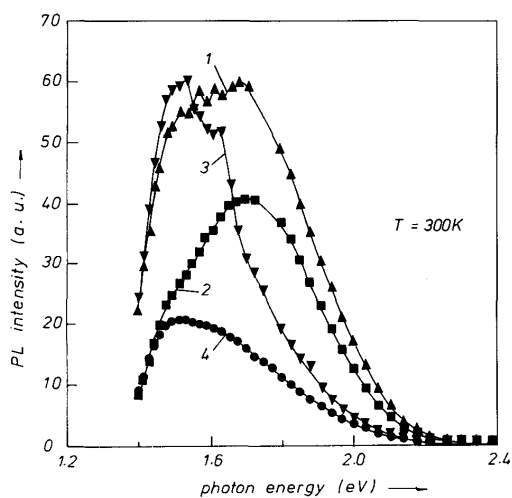

**Fig. 4.** PL spectra measured at RT on stabilized PS films (storage time of 4 weeks and then 3.5 months): excitation wavelength $\lambda$ = 352 nm, light power $P$ = 1 mW; detector: GaAs photomultiplier (–30°C; –1500 V), chopping frequency of 433 Hz [54]. Reproduced with permission from M. L. Ciurea, E. Pentia, A. Manea, A. Belu-Marian, I. Baltog, Visible photoluminescence in porous silicon prepared in different conditions - temperature dependence and decay, Phys. Stat. Sol. B 195, 637 (1996).

So, the films anodized for 90 min (curve 3) and those for 30 min (curve 4) have the main PL maximum positioned at 1.54 eV. Both kind of films were chemically etched under illumination (352 nm and power $P$ = 1 mW) for 2 min, stored firstly for 4 weeks and then for 3.5 months. The films anodized for 30 min, illuminated for 3 min, and similarly stored as previous films (4 weeks and then for 3.5 months) have the main maximum located at 1.72 eV, as curve 2 shows. Curve 1 is the envelope curve, showing both maxima. This film was stored for much longer time in both storage stages, namely for 16 weeks in the first one and 6.5 months in the second one. The origin of the two subbands was also clarified: we consider that the subband with maximum at 1.54 eV is due to QC effect/size effect as the anodization time for sample 3 is 3 times longer than for samples 2 and 4, while for the subband with maximum at 1.72 eV, the surface states are responsible.

**In conclusion**, we have shown experimentally, for the first time, that the broad band of PL in VIS contains two subbands only, positioned at 1.54 eV and 1.72 eV, and they are due to the QC effect and surface states, respectively.

Both nc-PS and films of Si NCs embedded in $SiO_2$ matrix are an important hope for fabrication of Si based optoelectronic devices. By finding this alternative solution, the III–V and II–VI materials (that are unprotective and unfriendly to the environment) on which at present optoelectronics is based, can be replaced with ecological Si – based materials. nc-PS has many applications in micro- and nanoelectronics, e.g. sacrificial layers, isolation walls etc., as mentioned before. We have fabricated a dielectrophoretic microfluidic device based on nanocrystalline porous silicon – polypyrrole bioelectrodes [70] designed to make a selection between leukocytes, i.e. to separate eosinophils, lymphocytes and monocytes.

## 3. Ge nanocrystals/nanoparticles embedded in oxide matrices ($SiO_2$, $TiO_2$) photosensitive in VIS-NIR-SWIR

### 3.1. Preparation

Films of Ge nanoparticles (NPs) / NCs in oxides ($SiO_2$, $TiO_2$) with high photoresponse at RT from VIS (400 – 700 nm) to NIR (700 nm – 1 μm) and SWIR (1 – 3 μm) up to 1325 nm were reported (Table 1) [71, 72].

**Table 1.** Films of Ge NPs/NCs in oxides ($SiO_2$, $TiO_2$) with high photosensitivity from VIS to NIR and SWIR up to 1325 nm: preparation conditions, structure and morphology.

| Film (Thickness) / Reference | MS deposition conditions | $T_{MS}$ (°C) | n-Si substrate | RTA conditions | Structure & morphology |
|---|---|---|---|---|---|
| Ge NPs-$SiO_2$ (200 nm) [72] | - targets: $SiO_2$ 262 W (7.3 nm/min), Ge 15 W (2 nm/min)<br>- 4 mTorr, 6N Ar | 300 | (100) 10–20 Ω·cm | – | - amorphous GeSiO film<br>- Ge/Si = 0.33 (at.%)<br>- Ge content is constant along film depth<br>- small Ge agglomeration near the interface with Si |
| Ge NPs-$SiO_2$ (123 nm) [72] | - targets: $SiO_2$ 262 W (7 nm/min), Ge 15 W (1.9 nm/min)<br>- 4 mTorr, 6N Ar | 400 | (100) 10–20 Ω·cm | – | N/A |
| *Ge NPs-SiO$_2$* (100 nm) [72] | - targets: SiO$_2$ 262 W (6.9 nm/min), Ge 15 W (1.7 nm/min) - 4 mTorr, 6N Ar | 500 | (100) 10–20 Ω·cm | – | - amorphous Ge NPs: 5–6 nm - few Ge NCs: 3.8±0.7 nm - almost whole amount of Ge is oxidized - Ge content decreases from film bottom towards free surface - thin Ge layer near the interface with Si: dense amorphous Ge NPs with more GeO$_x$ species than in the rest of film |
| *GeTiO$_2$* (180 nm) [71] | - targets: TiO$_2$ RF, Ge DC (60% vol. Ge) - 4 mTorr, 6N Ar | not heated | 400 nm SiO$_2$/Si | – | - amorphous layer |
| *Ge NCs-TiO$_2$* (190 nm) [71] | - targets: TiO$_2$ RF, Ge DC (60% vol. Ge) - 4 mTorr, 6N Ar | not heated | 400 nm SiO$_2$/Si | 550 °C 10 min Ar | - cubic Ge NCs: 4–5 nm size, 2–3 nm separation distance, high density ~4×10$^{18}$ cm$^{-3}$ - anatase TiO$_2$ NCs |
| *Ge NCs-TiO$_2$* (150 nm) [71] | - targets: TiO$_2$ RF, Ge DC (60% vol. Ge) - 4 mTorr, 6N Ar | not heated | 400 nm SiO$_2$/Si | 700 °C 10 min Ar | - very low Ge NCs density - big anatase TiO$_2$ NCs: 15–40 nm size (average 20 nm), separated by amorphous areas |

The films were deposited by magnetron sputtering (MS) at a pressure of 4 mTorr under 6N-purity Ar atmosphere. Ge NPs in SiO$_2$ films deposited at RT show very poor photoresponse, however high photosensitivity is achieved by heating the Si substrates during magnetron sputtering deposition. For this, Si substrates were heated at $T_{MS}$ = 300, 400 and 500 °C, and it was shown that by adjusting the deposition temperature, high values of photosensitivity including broadening of spectral response band (in SWIR), responsivity and internal quantum efficiency are achieved for Ge NPs-SiO$_2$ films (Table 2) [72]. $T_{MS}$ of 300 and 400 °C lead to high performance parameters, while in films deposited at 500 °C, Ge oxidation and loss are significant, thus leading to smaller cut-off wavelength in the photocurrent spectrum (e.g. 1267 nm in contrast to 1325 nm for $T_{MS}$ = 300 °C). The advantage of these photosensitive Ge NPs-SiO$_2$ films is the one-step fabrication consisting in the low temperature deposition without subsequent thermal treatment. So, the photosensitive Ge NPs-SiO$_2$ films were deposited by MS from two targets of Ge (15 W DC) and SiO$_2$ (262 W RF) on heated (100) n-Si substrates (10 – 20 Ω·cm resistivity), and they are formed of amorphous Ge NPs embedded in SiO$_2$ (Table 1). The Ge:SiO$_2$ vol. concentration is about 20:80 established from deposition rates obtained by ellipsometry. Sandwich *ITO/Ge NPs-SiO$_2$/n-Si/Al* structures were fabricated, the top ITO contact being deposited by MS, while the bottom Al is e-beam evaporated on the back side of Si wafer.

Photosensitive Ge NCs-TiO$_2$ films were prepared by MS deposition on Si substrates covered with 400 nm SiO$_2$ without heating them, followed by subsequent rapid thermal annealing (RTA). The deposition was done using two targets of Ge and TiO$_2$ leading to amorphous GeTiO$_2$ layers, while RTA was performed in Ar for 10 min at 550 and 700 °C for obtaining Ge-TiO$_2$ nanocrystalline films (Table 1). The best spectral photocurrent results were achieved for 550 °C RTA films formed of dense Ge NCs in *Ge NCs-TiO$_2$/SiO$_2$/n-Si* structures with thermally evaporated coplanar Al contacts [71]. RTA at higher temperature of 700 °C favors the Ge strong diffusion and its oxidation [73, 74], thus leading to very low Ge NCs density and consequently to the very poor photoconduction of the layer.

### 3.2. Morphology, structure and composition

The GeSiO films deposited on heated substrates ($T_{MS}$ = 300, 400 and 500 °C) present quite similar morphology (Table 1), meaning that they are amorphous (Figure 5b), being formed of amorphous Ge NPs in SiO$_2$ [72]. The film deposited at 300 °C has an atomic Ge:Si concentration ratio of 0.33 (energy dispersive X-ray spectroscopy (EDX) in Figure 5a in agreement with X-ray photoelectron spectroscopy (XPS)), the Ge content being constant along the film depth (200 nm). A small Ge agglomeration was observed near the interface with Si.

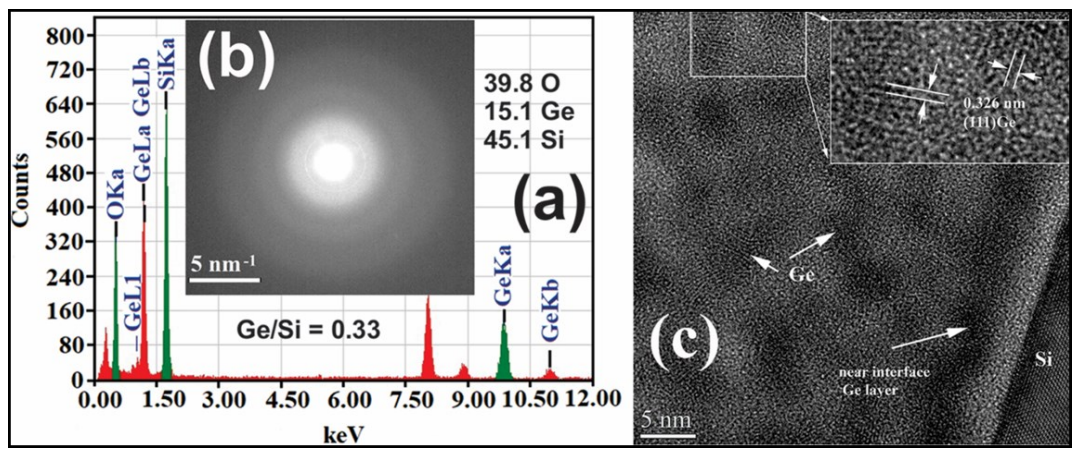

**Fig. 5.** Ge NPs-SiO$_2$ films deposited on Si substrates heated at 300 (a,b) and 500 °C (c): **a.** EDX analysis and **b.** selected area electron diffraction (SAED) pattern revealing a Ge:Si atomic concentration ratio of 0.33 in the amorphous GeSiO film; **c.** HRTEM image showing the amorphous Ge NPs (5 – 6 nm size) and few Ge NCs with 3.8 ± 0.7 nm size (inset), and the thin Ge layer with dense Ge NPs near the interface with Si substrate [72]. Reproduced with permission from I. Stavarache, V.S. Teodorescu, P. Prepelita, C. Logofatu, M. L. Ciurea, Ge nanoparticles in SiO$_2$ for near infrared photodetectors with high performance, Sci. Rep. 9, 10286 (2019). © 2019, Springer Nature.

Ge NPs-SiO$_2$ films deposited at 500 °C present besides amorphous Ge NPs (majority, 5 – 6 nm size), few Ge NCs with 3.8 ± 0.7 nm size (Figure 5c). Near the interface with Si, a thin Ge layer with dense amorphous Ge NPs forms, with more GeO$_x$ species than in the rest of film (XPS). This kind of interface layer was also observed in other GeSiO films, i.e. sol-gel films on Si substrate with 800 – 950 °C RTA [75] or sputtered films deposited at RT on quartz substrate, annealed in $H_2$ at 2 atm and 500 °C for 2h followed or not by a secondary annealing in $N_2$ at 1 atm and 800 °C for 2 h [76, 77].

XPS depth analysis of Ge NPS-$SiO_2$ films deposited at 500 °C showed that almost whole amount of Ge is oxidized and the Ge content decreases from the film bottom towards the free surface due to loss of Ge via GeO gas [72, 74, 78]. This agrees with the fact that the atomic Ge concentration is low in comparison to the case of films deposited at 300 °C.

The Ge NCs-$TiO_2$ films annealed by RTA at 550 °C are formed of cubic Ge NCs and anatase $TiO_2$ (Figure 6, Table 1). The Ge NCs have 4 – 5 nm size and 2 – 3 nm separation distance, and their density is high, of $\approx 4 \times 10^{18}$ cm$^{-3}$. The films annealed at 700 °C are formed of anatase $TiO_2$ NCs (15 – 40 nm size), but contain very low-density Ge NCs (Table 1) due to the strong diffusion and oxidation of Ge favored by the RTA high temperature [74].

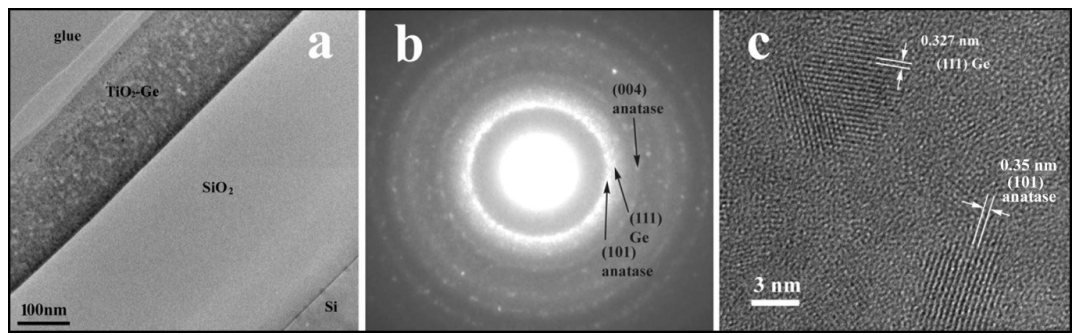

**Fig. 6.** 550 °C RTA Ge NCs-$TiO_2$ film on 400 nm $SiO_2$/Si substrate: **a.** cross-section TEM image – 180 nm film thickness, **b.** SAED pattern – cubic Ge and anatase $TiO_2$ spots and **c.** HRTEM image – Ge NCs (0.327 nm lattice fringe) and anatase $TiO_2$ [71]. Reproduced with permission from A.-M. Lepadatu, A. Slav, C. Palade, I. Dascalescu, M. Enculescu, S. Iftimie, S. Lazanu, V. S. Teodorescu, M. L. Ciurea, T. Stoica, Dense Ge nanocrystals embedded in $TiO_2$ with exponentially increased photoconduction by field effect, Sci. Rep. 8, 4898 (2018). © 2018, Springer Nature.

### 3.3. VIS-NIR-SWIR photosensing

It was shown that by adjusting the substrate temperature during MS deposition of Ge NPs-$SiO_2$ films, high values of photosensitivity, broadened spectral response band with cut-off wavelength extended in SWIR, high responsivity and internal quantum efficiency (*IQE*) over 100% are achieved for sandwich *ITO/Ge NPs-$SiO_2$/n-Si/Al* structures (Figure 7, Table 2) [72]. So, $T_{MS}$ = 300 and 400 °C lead to high performance parameters as 1325 nm cut-off wavelength in SWIR, 2.42 A/W responsivity and *IQE* = 445%, while in films deposited at 500 °C, Ge oxidation and loss are significant, thus leading to smaller cut-off wavelength in the photocurrent spectrum (e.g. 1267 nm in contrast to 1325 nm for $T_{MS}$ = 300 ºC).

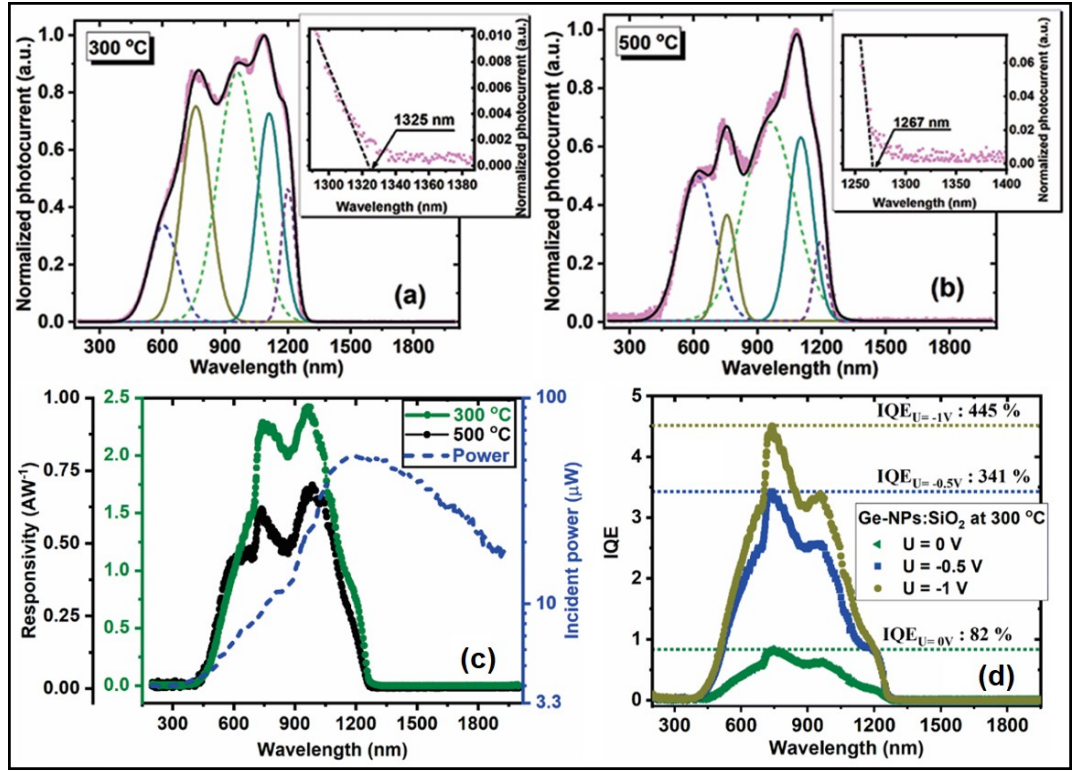

**Fig. 7.** Sandwich *ITO/Ge NPs-SiO$_2$/n-Si/Al* structures with Ge NPs-SiO$_2$ films deposited on Si substrates heated at 300 (a,b,c,d) and 500 ºC (b,c): **a.**, **b.** spectral photocurrent (−1V reverse bias, modulated light with 80 Hz frequency) measured at RT; **c.** spectral responsivity and incident power; **d.** spectral IQE for bias voltages of 0, –0.5 and –1V [72]. Reproduced with permission from I. Stavarache, V. S. Teodorescu, P. Prepelita, C. Logofatu, M.L. Ciurea, Ge nanoparticles in SiO$_2$ for near infrared photodetectors with high performance, Sci. Rep. 9, 10286 (2019). © 2019, Springer Nature.

The spectral photocurrents measured at RT and −1V reverse bias (under modulated light, 80 Hz frequency) on structures with Ge NPs-SiO$_2$ films deposited at $T_{MS}$ = 300 and 500 ºC are given in Figure 7a,b. One can see that they present broad spectral sensitivity interval from 400 to 1325 nm for $T_{MS}$ = 300 ºC, and from 400 to 1267 nm for $T_{MS}$ = 500 ºC (Table 2). Deconvolution of both spectra is possible by 5 maxima with similar positions, but with different intensities in function of the deposition temperature, i.e. a peak at 1200 nm due to the Ge-rich layer (dense Ge NPs) located near the interface with the Si substrate (Figure 5c), a peak at 1100 nm due to the Si substrate by capacitive coupling [71], and the remaining maxima and shoulders with positions bellow 1100 nm resulting from photo-effects in Ge NPs-SiO$_2$ films by Ge related states, significant part being located at Ge NP/SiO$_2$ interface.

**Table 2.** Photosensing parameters of Ge NCs/NPs-based structures: spectral sensitivity range, maximum values of responsivity and IQE at RT. Optical bandgap energies are also provided.

| Photosensitive structure / Reference | $T_{MS}$ (°C) | $T_{RTA}$ (°C) | $E_{g,opt}$ (eV) | Spectral sensitivity at RT (nm) | R (A/W) | IQE (%) |
|---|---|---|---|---|---|---|
| Sandwich ITO/Ge NPs-SiO$_2$/n-Si/Al [72] | 300 | – | 1.39 | 400 – **1325** | 2.42 | 445 |
| Sandwich ITO/Ge NPs-SiO$_2$/n-Si/Al [72] | 500 | – | 1.44 | 400 – **1267** | 0.69 | 118 |
| amorphous GeTiO$_2$/SiO$_2$/n-Si, Coplanar Al contacts [71] | – | – | 0.73 | 600 – **1200** | N/A | N/A |
| Ge NCs-TiO$_2$/SiO$_2$/n-Si, Coplanar Al contacts [71] | – | 550 | 1.14 | 600 – **1250** | N/A | N/A |

The responsivity spectra (ratio between the spectral photocurrent and spectral incident optical power), $R_\lambda = I_{ph}(\lambda)/P_{inc}(\lambda)$ from Figure 7c show a high maximum value of 2.42 A/W, that was achieved for samples with $T_{MS}$ = 300 °C. The samples with higher deposition temperature $T_{MS}$ = 500 °C present a lower responsivity of 0.69 A/W.

The internal quantum efficiency *IQE* was calculated using the formula $IQE = [hc/q\lambda(1-R)] \times R_\lambda$, and *IQE* spectra obtained for bias voltages of 0, –0.5 and –1V are shown for $T_{MS}$ = 300 °C in Figure 7d. IQE maximum values of 445%, 341% and 82% are obtained for –1V, –0.5 and 0 V bias voltage respectively, while for $T_{MS}$ = 500 °C lower values are obtained (118%, 89% and 21% for the same biases). So, we obtained high IQE with values exceeding 100%.

The high photosensitivity, dependent on the electric field is explained by the increased lifetime of electrons participating in the photoconductivity as result of trapping of holes on Ge related defects/traps [72].

The advantage of these photosensitive Ge NPs-SiO$_2$ films is the one-step fabrication consisting in the low temperature deposition without subsequent thermal treatment.

We have to remark that Ge-SiO$_2$ films deposited at RT have very poor photosensitivity [72]. We showed that sol-gel films of amorphous Ge NPs in SiO$_2$, some NPs containing tetragonal lattice traces, present broad sensitivity from 350 to 900 nm [75, 79]. In this case, the photosensitivity is due to localized states from Ge NPs/matrix interface and tetragonal Ge clusters. The enhancement of photocurrent is also achieved as by trapping carriers of one type, the lifetime of free carriers with opposite sign is increased.

The photocurrent spectra measured (at RT, 10 V bias voltage, 120 Hz frequency chopped light) on the as-deposited amorphous *GeTiO$_2$/SiO$_2$/n-Si* and RTA annealed *Ge NCs-TiO$_2$/SiO$_2$/n-Si* structures with coplanar Al electrodes geometry (Figure 8b) are presented in Figure 8a. One can see that the spectral sensitivity extends from 600 nm up to 1250 nm for Ge NCs-TiO$_2$ structures, and up to 1200 nm for amorphous as-deposited structures. Both spectra present two main peaks, one broad at 870 nm for 550 °C RTA or 910 nm for as-deposited structure, and the other narrow peak at the same position of 1100 nm. The broad peak at 870 nm (550 °C RTA) or 910 nm (as-deposited) is attributed to the photo-effects in Ge NCs-TiO$_2$ and amorphous GeTiO$_2$ layers respectively, while the 1100 nm peak is due to Si substrate by the surface photovoltage and gating effects [80, 81]. The photocurrent of 550 °C RTA Ge NCs-TiO$_2$ structures is about 10 times higher than for as-deposited films, the increase being correlated with the formation of Ge NCs. For wavelengths of about 1100 nm, the spectral photocurrent has important contribution from Si substrate by surface photovoltage and gating effect. By increasing RTA temperature to 700 °C, photocurrent spectrum shows only the 1100 nm peak due to Si substrate, as the Ge NCs density is diminished and therefore the film transparency becomes high.

The voltage dependence of the intensities of the 870 and 1100 nm main peaks (Figure 8c) was calculated by using the values from the deconvolution of photocurrent spectra measured for different bias voltages (from 7 to 20 V, 1V step) on the 550 °C RTA structure (Figure 8c). The deconvolution was carried out by considering three peaks centered at 1100, 870 and 570 nm for the 600 – 1250 nm interval. It is obtained that for low voltages up to 12 V, the photocurrent increases exponentially with the bias voltage (Figure 8d) being due to the electrostatic doping by the field effect. This means that the variation of carrier concentration by external field leads to a hole depletion zone in the Ge-TiO$_2$ layer induced by field effect by charging regions at both Al/Ge-TiO$_2$ and Ge-TiO$_2$/SiO$_2$ interfaces (Figure 8b). The hole depletion zone is close to the region of negatively biased contact with major contribution to the photocurrent. Si has a beneficial role as the photoconduction on the layer increases by electrostatic doping effect.

Further improvements of Ge NPs/NCs photosensitivity were achieved by our team by following different directions, i.e. the most important one being the alloying of Ge with little Si in order to hinder the Ge fast diffusion and thus for improving the thermal stability of NCs [82].

Furthermore, by using nanocrystallized HfO$_2$ matrix instead of SiO$_2$ and TiO$_2$, high quality Ge-rich SiGe NC/matrix interface is ensured. Thus, we obtained strained Ge-rich SiGe NCs embedded in nanocrystallized HfO$_2$ layers with photoelectric sensitivity extended in SWIR up to 1800 nm at RT or even 2000 nm for cooled structures (100 K) [83].

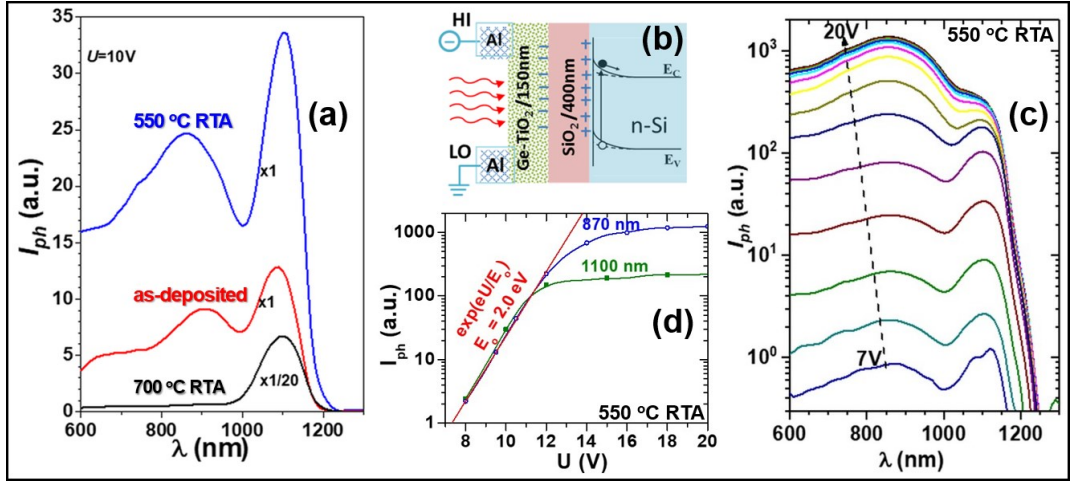

**Fig. 8.** Photosensitive *Ge NCs-TiO$_2$/SiO$_2$/n-Si* structures (coplanar Al contacts), RTA annealed at 550 (a,b,c,d) and 700 °C (a) and as-deposited *GeTiO$_2$/SiO$_2$/n-Si* structures (a): **a.** spectral photocurrent curves measured for 10 V bias voltage on as-deposited and RTA annealed structures; **b.** schematic of the structure with coplanar Al contacts for photoelectrical measurements; **c.** photocurrent spectra measured for different bias voltages (from 7 to 20 V, 1V step) on 550 °C RTA structure with main peaks of 870 and 1100 nm; **d.** voltage dependence of the main peaks intensities from Figure 8c [71]. Spectral photocurrent was measured at RT under illumination with chopped light (120 Hz frequency). Reproduced with permission from A.-M. Lepadatu, A. Slav, C. Palade, I. Dascalescu, M. Enculescu, S. Iftimie, S. Lazanu, V. S. Teodorescu, M. L. Ciurea, T. Stoica, Dense Ge nanocrystals embedded in TiO$_2$ with exponentially increased photoconduction by field effect, Sci. Rep. 8, 4898 (2018). © 2018, Springer Nature.

In this case, the high sensitivity is due to the high-quality of interface between SiGe NCs and nanocrystallized HfO$_2$ matrix. Hydrogen plasma annealing is also beneficial for enhancing the photoresponse [84]. High photosensitivity in SWIR was obtained by alloying Ge with Sn, for this kind of structures a cut-off up to 2500 nm was achieved [85, 86]. Of course, the obtaining technology is more difficult as the Ge and Sn miscibility is very low (~1% in bulk).

Additionally, DFT calculations for mainly determining the bandgap energy were performed, that are useful for design and characterization of Ge and SiGe NCs-based photosensitive structures [87].

## 4. Ge nanocrystals/quantum dots/nanoparticles embedded in classical SiO$_2$ and high-κ oxides (HfO$_2$, Al$_2$O$_3$) as floating gate in non-volatile memories

### 4.1. Fabrication of 3-layer floating gate memory structures

Wafers of p-type Si (100) with 7 – 14 Ω·cm resistivity were used as substrates for fabrication of MOS-like capacitors. The cleaning of Si wafers follows standard procedures as detailed in [88]. The initial structures (before post-annealing) with 3-layer stack of *gate oxide/intermediate layer of Ge or Ge-oxide/tunnel oxide layer or 2-layer stack/p-Si* represent the starting point for obtaining floating gate NVMs. As oxides, classical $SiO_2$ and high-κ dielectrics $HfO_2$ and $Al_2O_3$ we used. $SiO_2$ has the advantage of the near-perfect $SiO_2$/Si interface ensuring low interface states density ($D_{it}$). Also, $SiO_2$ has a very low electronic defect density. However, downscaling of devices size requires replacement of $SiO_2$ with high-κ dielectrics principally due to the leakage by detrimental tunneling through the very thin $SiO_2$ into the gate electrode in the CMOS transistor [89]. By using thicker high-κ dielectrics, the same capacitance is kept and the leakage current due to the direct tunneling is lowered at the same time [90]. The fully-CMOS compatible $HfO_2$ ($\kappa \approx 20$) is already greatly used in microelectronics, as gate dielectric in high-κ metal gate MOSFETs [91]. The high-κ $Al_2O_3$ ($\kappa \approx 10$) is also interesting for microelectronic applications, offering the same benefits as the high-κ $HfO_2$, i.e. enhanced data retention and lower operation voltage, and thus being suitable for small devices and high density NVMs [92]. One problem related to high-κ oxides is that they have higher electronic defect density because their chemical bonds cannot easily relax, in contrast to $SiO_2$ with low coordination number that favors the healing/repairing of dangling bonds [93]. However, the quality of high-κ oxide layers regarding to electronic defects density can be improved by conducting deposition and annealing treatments under controlled conditions.

So, the *gate oxide/ intermediate layer of Ge or Ge-oxide/ tunnel oxide layer or 2-layer stack/ p-Si* structures to be obtained as starting point were *$SiO_2$/ Ge/ $SiO_2$/ p-Si* (30/ 3 – 5/ 5 – 7 nm layer thickness) [94], *$HfO_2$/ Ge/ $HfO_2$/ p-Si* (20 – 30/ 5 – 10/ 5 – 10 nm) [95], *$HfO_2$/ Ge-$HfO_2$/ $HfO_2$/ p-Si* (20 – 50/ 8/ 5 – 10 nm) [96] and *$Al_2O_3$/ Ge/ [$Al_2O_3$/$SiO_2$]/ p-Si* (10/ 15/ [4/3] nm) [88]. The thickness intervals for all layers, i.e. gate and tunnel oxide and floating gate result from the optimization process using the feedback from the memory properties measurements. The tunnel region is either an oxide layer ($SiO_2$ or $HfO_2$) or a 2-layer stack of 2 different oxides such as $Al_2O_3$/$SiO_2$ on Si. The tunnel $SiO_2$ (5 – 7 nm thickness for single tunnel layer; 3 nm in the tunnel 2-layer stack) is of very high quality, being obtained by rapid thermal oxidation (at 1050 ºC). Tunnel layers of $HfO_2$ (5 – 10 nm thickness) and $Al_2O_3$ (4 nm) are obtained by MS deposition. The rest of layers, i.e. the intermediate layer of either continuous Ge (5 – 10 nm) or co-deposited Ge-oxide (8 nm Ge-$HfO_2$ with 70:30 vol.% composition) and the gate oxide layer (30 nm $SiO_2$, 20 – 50 nm $HfO_2$ and 10 nm $Al_2O_3$) are also deposited by MS. Sputtering deposition is performed at 4 mTorr working pressure using 6N-purity Ar, without heating the substrates. The base pressure in the deposition chamber reaches at least $10^{-7}$ Torr. The intermediate Ge and gate $SiO_2$ layers in the *$SiO_2$/ Ge/ $SiO_2$/ p-Si* structures were deposited by using powers of 10 W DC for Ge target, and 120 W RF for $SiO_2$ one. In the case of *$HfO_2$/ Ge/ $HfO_2$/ p-Si* and *$HfO_2$/ Ge-$HfO_2$/ $HfO_2$/ p-Si* structures, we used powers of 40 – 60 W RF for $HfO_2$, and 5 – 10 W DC for Ge. For *Al$_2$O$_3$/ Ge/ [Al$_2$O$_3$/SiO$_2$]/ p-Si* structures, 100 W RF was applied on the Al$_2$O$_3$ target, and 30 W RF on the Ge target.

After deposition, RTA at atmospheric pressure in a rapid thermal processor, under Ar or N$_2$ atmospheres (6N purity) was performed on the 3-layer structures for Ge nanostructuring. RTA processing ensures the formation of Ge NCs with controlled size, density and location in the floating gate, positioned at tunnelable distance from the Si substrate in the 3-layer. This is valid for both as-deposited Ge or Ge-oxide intermediate layer [73]. The benefit from RTA is that Ge crystallization is faster than Ge diffusion in contrast to furnace annealing. In this way, dense Ge NCs layers were obtained. So, *SiO$_2$/ Ge/ SiO$_2$/ p-Si* were RTA annealed at 900 °C in Ar for 15 min, and the floating gate of Ge NCs acting as charge storage centers, being embedded in the amorphous SiO$_2$ matrix was obtained. In the case of HfO$_2$-based structures, different RTA temperatures from 600 to 850 °C were studied. It was shown that for NVMs with Ge NCs/QDs-HfO$_2$ floating gate (22 nm thick gate HfO$_2$ layer), the optimum RTA temperature is 600 °C for which the best memory properties are obtained, together with high-density Ge NCs/QDs that are the main contributors to the charge storage. These structures were RTA annealed under 6N-purity N$_2$ for 8 min. We showed that the RTA duration is correlated with the temperature (from 600 °C for 8 min, 650 °C for 8 min and 4 min, respectively to 700 °C for 4 min) in connection with the competition between the strongly temperature-dependent processes of Ge segregation and diffusion [96]. The increase of RTA temperature leads to the decrease of Ge QDs density due to the enhanced Ge fast diffusion into HfO$_2$. Additionally, RTA also produces the nanocrystallization of HfO$_2$. For *Al$_2$O$_3$/ Ge/ [Al$_2$O$_3$/SiO$_2$]/ p-Si* structures, the situation is much different as the performance of Al$_2$O$_3$ based floating gate NVMs is enhanced by the formation of additional Ge-related storage centers as Ge-related local electronic states in addition to Ge-rich NCs. RTA in the temperature range of 550 – 900 °C was performed for 1 min in N$_2$. The best memory structures were obtained for RTA at 600 and 650 °C for which Ge remains amorphous and strongly diffuses into the Al$_2$O$_3$ layers. Samples annealed at 700 °C showed also very good memory properties (high memory window and very good retention) that corresponds to the state of the art of NCs based floating gate NVMs. For these samples (700 °C RTA), Ge NCs in the intermediate layer and amorphous Ge NPs in the tunnel Al$_2$O$_3$ layer were observed. RTA at higher temperatures (800 and 900 °C) leads to the strong decrease of the density of Ge-related states, and to a smaller density of charge storage centers (NCs and Ge-related states), and consequently to the memory window decrease.

One particular case related to HfO$_2$ structures is given by 3-layers with thick gate oxide (64 nm thickness in contrast to 22 nm gate HfO$_2$ layer) [97]. *HfO$_2$/ Ge-HfO$_2$ (65:35 vol.% Ge: HfO$_2$)/ HfO$_2$/ p-Si* structures were also deposited by MS, the designed thicknesses being 60 – 65/ 6 – 8/ 10 – 14 nm. Subsequent RTA was performed at 620 ºC in N$_2$, leading to the formation of amorphous Ge NPs arranged in two rows inside HfO$_2$ matrix in the floating gate. Additionally, HfO$_2$ has orthorhombic/tetragonal structure in the floating gate and in the thin regions adjacent to it inside tunnel and control top HfO$_2$ (along ≈ 5 nm thickness of each oxide layer), thus being ferroelectric. This leads to the cumulative contribution of Ge NPs charge storage centers and ferroelectric HfO$_2$ matrix to the memory effect.

The HfO$_2$ and Al$_2$O$_3$-based NVM structures with MOS-like configuration were completed by depositing top and bottom Al contacts by thermal evaporation. In the case of SiO$_2$-based structures with Ge NCs in SiO$_2$, memory devices with cross-bar configuration formed of 5568 (100 × 100 μm$^2$ cell size) and 871 (300 × 300 μm$^2$ cell size) memory cells were fabricated as described in [94].

### 4.2. Morphology, structure and composition

All as-deposited 3-layer structures with not heated Si substrates during MS deposition are amorphous. By RTA, Ge nanostructuring is achieved. The morphology and structure of 3-layer stacks is tailored by conducting RTA under different conditions of temperature and duration, and then charge storage properties were investigated.

HRTEM images evidenced the formation of Ge NCs in the amorphous SiO$_2$ matrix in the floating gate of the *SiO$_2$/ Ge/ SiO$_2$/ p-Si* (30/ 3 – 5/ 5 – 7 nm) structures RTA annealed at 900 ºC [94]. The Ge NCs are the charge storage centers inside the floating gate.

In the case of *HfO$_2$/ Ge/ HfO$_2$/ p-Si* (20 – 30/ 5 – 10/ 5 – 10 nm) structures, the RTA temperature for the formation of Ge NCs is much lower, the optimum temperature being around 600 ºC [95]. It was shown that the 600 ºC RTA leads to the formation of high density Ge NCs with 5–7 nm size in the floating gate, the Ge NCs acting as the charge storage centers (Figure 9a,b). Depth-analysis XPS evidenced a 60/40 ratio of the intensities of metallic and oxidized Ge peaks positioned similarly to those for as-deposited Ge layer (Figure 9c). Additionally, HfO$_2$ has monoclinic structure. The RTA at higher temperature of 850 ºC drastically changes the 3-layer morphology, making it less visible than for 600 ºC RTA. So, HfO$_2$ is crystallized all over structure thickness, and Ge spreads into the crystallized HfO$_2$ matrix. The high temperature RTA favors the fast Ge diffusion against Ge segregation, thus hindering the formation of Ge NCs. This leads to a strongly decreased Ge NCs density and consequently to the narrowing of memory window. The RTA duration is correlated with the temperature in connection with the competition between the strongly temperature-dependent processes of Ge segregation and diffusion [96].

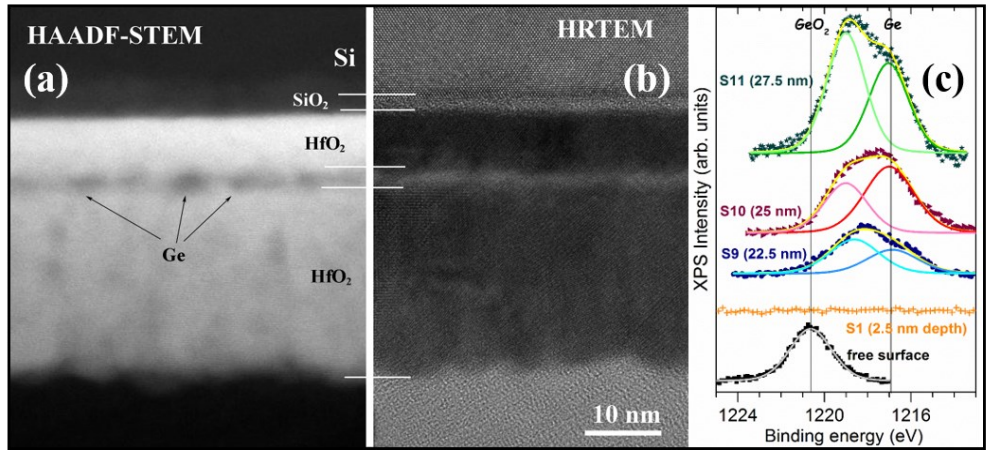

**Fig. 9.** *HfO$_2$/ Ge/ HfO$_2$/ p-Si* NVMs with 600 ºC RTA: **a.**, **b.** High-angle annular dark field (HAADF-STEM) and HRTEM cross-section images evidencing the formation of Ge NCs (arrows) in the floating gate; **c.** XPS spectra and deconvolution curves for the Ge 2p3/2 core level (XPS depth ≈ 2.2 nm) at the free surface and at the depths of 2.5, 22.5, 25 and 27.5 nm (corresponding to 2.5 nm Ar+ sputtering steps 1, 9, 10 and 11 respectively) [95]. Reprinted from Scripta Mater. 113, 135 (2016), A. Slav, C. Palade, A. M. Lepadatu, M. L. Ciurea, V. S. Teodorescu, S. Lazanu, A. V. Maraloiu, C. Logofatu, M. Braic, A. Kiss, How morphology determines the charge storage properties of Ge nanocrystals in HfO$_2$, Copyright (2016), with permission from Elsevier.

By using *HfO$_2$/ Ge-HfO$_2$/ HfO$_2$/ p-Si* (20 – 25/ 8/ 5 – 10 nm) structures as starting point [96], i.e. by using the Ge-HfO$_2$ co-deposition approach for deposition of intermediate layer instead of depositing a continuous Ge layer, a floating gate with well-defined single layer of Ge QDs in HfO$_2$ is obtained by 600 ºC RTA. The Ge QDs have 2 – 3 nm size and $4 – 5 \times 10^{11}$ QDs/cm$^2$ density, and are located at the crossing of HfO$_2$ NCs boundaries (Figure 10a,b,c). Inside the floating gate, the HfO$_2$ NCs have tetragonal/orthorhombic lattice stabilized by the presence of Ge atoms (as result of Ge diffusion that is strong even for relatively low temperatures), and 8 nm diameter that represents the separation distance between the Ge QDs. In this way, we obtained in the floating gate a single layer of Ge QDs, well separated and unstacked. This means a good lateral separation of the Ge QDs storage centers that will hinder the lateral charge loss improving the charge retention and therefore increasing the memory performance, i.e. the abrupt low to high transition (from the inversion to accumulation) in the *C – V* hysteresis curve. The Ge QDs have hexagonal crystalline structure (very close lattice parameters with HfO$_2$ structure) by topotactic crystallization on the tetragonal/ orthorhombic HfO$_2$ structure (similar to epitaxial growth), the structure of HfO$_2$ NC acting as a trigger for the Ge QDs lattice formation. Depth-analysis XPS shows a very high ratio of 89/11 of peak intensities of metallic and oxidized Ge in the floating gate that keeps the position corresponding to as-deposited Ge layer (Figure 10d), thus evidencing the sharp interfaces of floating gate with gate and tunnel $HfO_2$ layers. The 89/11 ratio for the Ge-$HfO_2$ co-deposition approach is much higher than the one of 60/40 obtained for the continuous Ge intermediate layer approach. The single layer arrangement and good lateral separation of Ge QDs together with the sharp floating gate / adjacent oxide layer interfaces will lead to an enhanced memory performance in comparison to *$HfO_2$/ Ge/ $HfO_2$/ p-Si* NVMs. Some Ge diffuses in gate $HfO_2$ in which $HfO_2$ NCs are big (up to 20–30 nm) and have monoclinic structure.

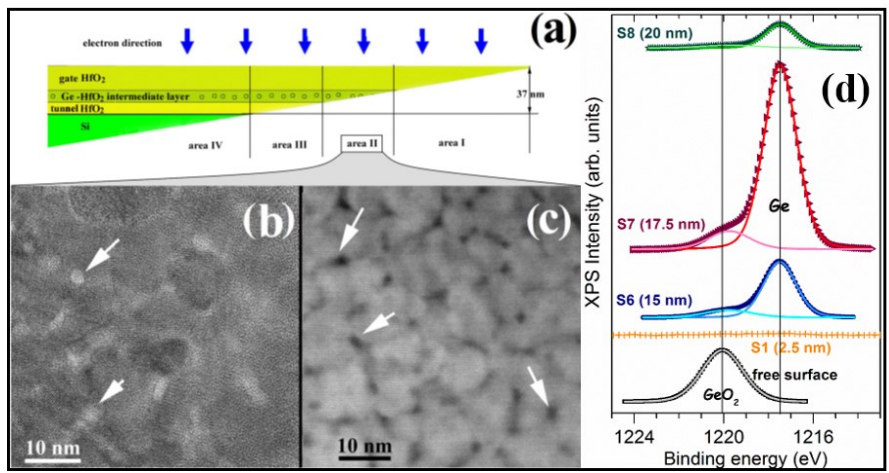

**Fig. 10.** *$HfO_2$/ Ge-$HfO_2$/ $HfO_2$/ p-Si* NVMs (22/ 7 – 8/ 8 nm) with 600 ºC RTA: **a.**, **b.**, **c.** plan view imaging schematic, and HRTEM and similar HAADF-STEM Z contrast mode images from area II with arrowed Ge QDs; **d.** XPS spectra and deconvolution curves for the Ge 2p3/2 core level at the free surface and at the depths of 2.5, 15, 17.5 and 20 nm [96]. © IOP Publishing. Reproduced with permission. All rights reserved.

The particular case of *$HfO_2$/ Ge-$HfO_2$/ $HfO_2$/ p-Si* structures (60 – 65/ 6 – 8/ 10 – 14 nm) with thick top $HfO_2$ layer shows the formation of a completely different morphology by 620 ºC RTA [97]. The structure practically changes into a multilayer structure with layers sequence of *m-$HfO_2$/ o/t-$HfO_2$/ two rows of a-Ge NPs in o/t-$HfO_2$/ o/t-$HfO_2$/ m-$HfO_2$/ p-Si*, where m-, o/t- and a- refer to monoclinic, o/t- orthorhombic/tetragonal and a- amorphous structures. Inside floating gate, amorphous spheroidal Ge NPs with 2 – 3 nm size are formed, being positioned in two rows (Figure 11a,b). Each row of Ge NPs is located near the interfaces of the floating gate with control and tunnel $HfO_2$ layers respectively. Few small Ge clusters with hexagonal structure are also formed. Ge was not detected in the control and tunnel $HfO_2$ layers. This specific morphology, i.e. arrangement of amorphous Ge NPs into two rows is the result of strain constrictions induced by the thick control $HfO_2$ layer (60 – 65 nm thickness compared to 20 – 25 nm thick gate oxide layer). Strain relaxation is achieved by Ge diffusion toward the borders of the floating gate layer. Additionally, $HfO_2$ NCs (2 – 3 nm) have orthorhombic/tetragonal structure in the floating gate, and also in the adjacent thin areas inside control and tunnel oxide layers due to the strain induced by the positioning of Ge near them. So, the $HfO_2$ matrix crystallizes first, i.e. in orthorhombic phase in the neighborhood of amorphous Ge content zone. In the remaining control and tunnel $HfO_2$ layers areas, $HfO_2$ NCs (10 – 30 nm) are monoclinic. Depth-analysis XPS evidenced that in both rows of Ge NPs Ge is mainly in metallic state, i.e. the ratio of the intensities of metallic and oxidized Ge peaks is about 80/20 (Figure 11c). So, in this kind of structures, the amorphous Ge NPs and the hexagonal Ge clusters are the charge storage centers. To the contribution of the Ge-related centers to the memory effect will cumulatively add the contribution of ferroelectric orthorhombic $HfO_2$ present in the floating gate and in the adjacent thin areas from control and tunnel layer. On the other hand, the bilayer Ge NPs floating gate morphology works to the benefit of improved memory properties (charge storage and retention time) compared to the single layer floating gate [98, 99].

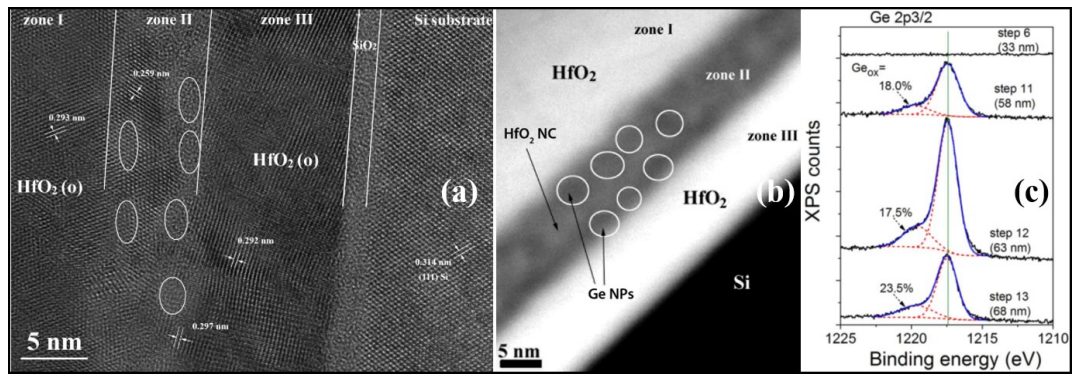

**Fig. 11.** *$HfO_2$/ Ge-$HfO_2$/ $HfO_2$/ p-Si* NVMs (64/ 7 – 7.5/ 11 nm) RTA annealed at 620 °C: **a.**, **b.** HRTEM and HAADF-STEM cross-section images revealing the two rows of Ge NPs (by ellipses) located near the interfaces of the floating gate with control and tunnel $HfO_2$ layers; **c.** XPS spectra and deconvolution curves for the Ge 2p3/2 core level at the depths of 33, 58, 63 and 68 nm [97]. © IOP Publishing. Reproduced with permission. All rights reserved.

In comparison to the NVM structures using oxides of only $HfO_2$ or $SiO_2$, the structures with $Al_2O_3$ for gate oxide and $Al_2O_3/SiO_2$ stack for tunnel oxide show different structure and morphology with more complex changes by RTA at temperatures from 550 to 900 °C as the lateral Ge diffusion into the $Al_2O_3$ layers is very strong even for 600 °C RTA [88]. So, the *$Al_2O_3$/ Ge/ [$Al_2O_3/SiO_2$]/ p-Si* structures annealed at 600 °C are amorphous. More than that, the strong diffusion of Ge into the adjacent gate and tunnel $Al_2O_3$ layers is evidenced, i.e. 25.% Ge to 77.5% Al in the gate $Al_2O_3$ layer. 650 °C RTA results are similar to 600 °C RTA.

By RTA at 700 °C RTA, amorphous mixed (Al oxide and Ge) Ge-rich NPs are formed in the tunnel $Al_2O_3$ layer (Figure 12a,b,d). In the floating gate, there are some crystallized Ge areas corresponding to 2 – 3 nm Ge NCs with cubic structure (Figure 12a,c,d). The lateral Ge diffusion is stronger, and voids and low-density amorphous zones are present in the intermediate Ge layer (also for 650 °C RTA), while the majority of Ge segregates at the interface with $Al_2O_3$ layers or inside the tunnel $Al_2O_3$ layer. Ge diffusion and nanocrystallization are affected by the strain induced by nanocrystallization during annealing and by the strain induced by the gate $Al_2O_3$ layer on top of intermediate Ge layer.

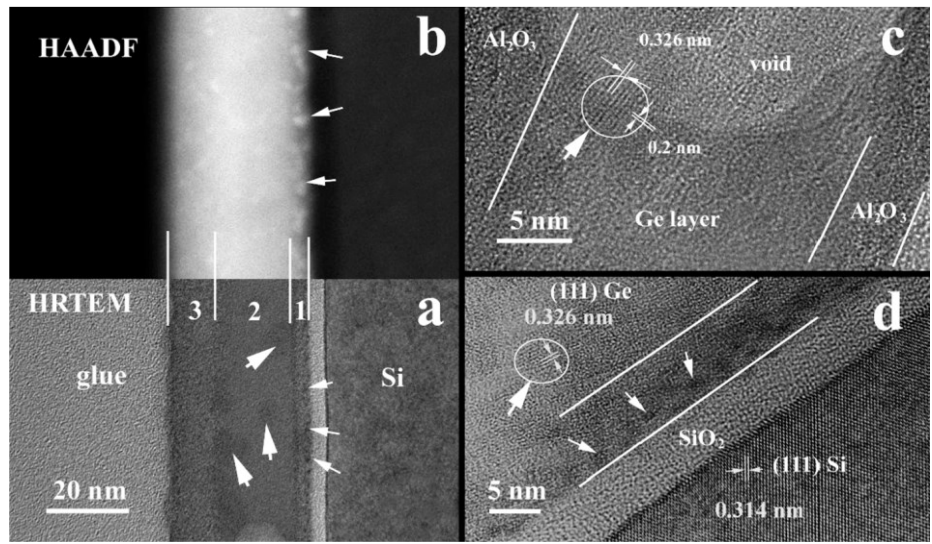

**Fig. 12.** *$Al_2O_3$/ Ge/ [$Al_2O_3$/$SiO_2$]/ p-Si* NVMs RTA annealed at 700 °C: **a.**, **b.** HRTEM and HAADF images with the same magnification (cross-section); **c.**, **d.** high resolution details. Big arrows indicate the formation of Ge NCs (cubic structure by 0.326 nm lattice fringe) in the intermediate Ge layer (images **a**, **c**, **d**), while small arrows indicate the formation of amorphous NPs inside the tunnel $Al_2O_3$ layer (images **a**, **b**, **d**) [88]. Reprinted from Appl. Surf. Sci. 542, 148702 (2021), I. Stavarache, O. Cojocaru, V. A. Maraloiu, V. S. Teodorescu, T. Stoica, M. L. Ciurea, Effects of Ge-related storage centers formation in $Al_2O_3$ enhancing the performance of floating gate memories, Copyright (2021), with permission from Elsevier.

Thus, if the top part of the structure has more freedom in strain relaxation, diffusion and nanocrystallization, the bottom part experiences more stress resulting in less structural transformation freedom. Strain relaxation will be achieved by Ge diffusion toward the borders of the floating gate layer.

By increasing RTA temperature to 800 °C RTA, the 3-layer morphology is altered (Ge NCs are no longer formed in the floating gate, NCs of mixed Al oxide and Ge rich in Ge are formed near the interface with tunnel $Al_2O_3$ layer, lateral Ge diffusion is stronger, and also voids are more visible in the Ge layer compared to 700 °C RTA.

Inter-diffusion, chemical reactions and intermixing processes at the Ge/$Al_2O_3$ interface are studied and reported in literature [100]. One factor that could influence the stability of $Al_2O_3$ in presence of Ge at high temperatures could be the non-stoichiometry of $Al_2O_3$ deposited by sputtering in Ar inert atmosphere, being usually obtained with a slight oxygen deficiency.

The NVMs with $Al_2O_3$ benefit from the high density of Ge related states acting (NCs and local electronic states by deep energy levels present in the amorphous structure) as charge storage centers, enhancing their performance. Less degradation of the resistivity of the tunnel oxide layer is expected due to the formation of mixed $AlGeO_2$ oxide.

### 4.3. Charge storage properties

The operation mechanism of 3-layer NVM capacitors with floating gate of Ge QDs in $HfO_2$ is given in Figure 13 [96].

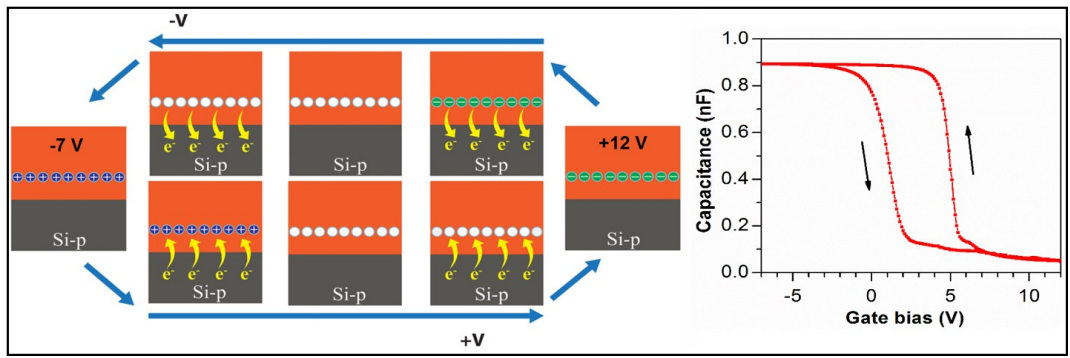

**Fig. 13.** Schematic of floating gate NVM working mechanism exemplified for 3-layer memory capacitors with floating gate of Ge QDs in $HfO_2$ together with the $C - V$ hysteresis loop. The transfer of the charges from/to Ge QDs controlled by the bias is evidenced. Arrows added to the $C - V$ hysteresis curve show the direction of the voltage programming cycle to understand the electron or hole accumulation/depletion mechanism [96]. © IOP Publishing. Reproduced with permission. All rights reserved.

This mechanism is applicable to any NVM with Ge NCs/QDs/NPs embedded in oxide ($SiO_2$, $Al_2O_3$). In general, the flatband voltage shift is caused by the charge transfer between Si substrate and Ge QDs in floating gate, and Si/$HfO_2$ interface states as well. The tunneling through the tunnel oxide takes place by intermediate states irrespective of 3 and 10 nm tunnel oxide thickness, the thickness being selected from optimization of memory properties. So, the charge transfer between Si substrate and Ge QDs/NCs/NPs is ensured. The transfer of charges from/to Ge QDs controlled by the applied voltage / bias (schematic in Figure 13) is important and results in changes of the flatband voltage, and consequently in hysteresis loops ($C - V$ loop in Figure 13). The memory window $\Delta V_{FB}$ is defined as the difference between flatband voltages $V_{FB}$ corresponding to the two hysteresis branches [97]. The flatband voltage depends on the charging state of the Ge QDs. In our case of p-type Si substrate, the shift of the flatband voltage to positive voltages is obtained by the negative charging of Ge QDs together with possible negative charge at Si/HfO$_2$ interface. The best memory properties of NVMs using HfO$_2$ were obtained for the structures *HfO$_2$/ Ge-HfO$_2$/ HfO$_2$/ p-Si* NVMs (22/ 7 – 8/ 8 nm) annealed by RTA at 600 ºC and *HfO$_2$/ Ge-HfO$_2$/ HfO$_2$/ p-Si* NVMs (64/ 7 – 7.5/ 11 nm) with RTA at 620 ºC (Figure 14, Table 3).

One can see that the 600 ºC RTA *HfO$_2$/ Ge-HfO$_2$/ HfO$_2$/ p-Si* NVMs (22/ 7 – 8/ 8 nm) structures present $C - V$ hysteresis loops with high memory window $\Delta V = 3.8 \pm 0.5$ V due to charge storage in Ge QDs arranged in the single layer [96]. The memory window changes very slightly (4%) with the frequency variation (100 kHz to 1 MHz) due to Ge/GeO$_x$ interface and/or series resistance.

As a result of the good lateral separation of Ge QDs (by the HfO$_2$ NCs at the crossing of their boundaries), the charge retention curve ($C - t$) shows a sharp/fast decrease of capacitance $C$ of only 14% in the first 3000 – 4000 s followed by a slow decay. The low lateral tunneling (lateral charge loss is hindered) ensured by the good lateral separation of Ge QDs is demonstrated by the small fast capacitance decrease. By extrapolating the $C - t$ curve to 10 years, the capacity reaches 50% of its initial value. The *HfO$_2$/ Ge-HfO$_2$/ HfO$_2$/ p-Si* NVMs (64/ 7 – 7.5/ 11 nm) with RTA at 620 ºC show high memory windows of up to 6.1 V, while the $C - t$ curve shows a capacitance decrease with 20% after 6000 s similar to *HfO$_2$/ Ge-HfO$_2$/ HfO$_2$/ p-Si* NVMs (22/ 7 – 8/ 8 nm) structures (i.e. 16% decrease of $C$). The high memory window is due to both contributions of Ge-related storage centers (amorphous Ge NPs and hexagonal Ge clusters) and ferroelectric orthorhombic HfO$_2$ (in floating gate and in adjacent thin areas from control and tunnel layers) [97, 101]. Regarding charge storage in Ge NPs and hexagonal Ge clusters, high densities of up to $6.3 \times 10^{12}$ stored electrons/cm$^2$ were achieved [97]. Additionally, the bilayer Ge NPs floating gate morphology favors the obtaining of improved memory properties compared to the single layer floating gate.

At present, we are dealing with the fabrication of multilayered floating gate NVM devices with charge storage nodes of Ge$_x$Si$_{1-x}$ NCs embedded in nanocrystallized HfO$_2$ [102]. Such kind of NVMs benefit from the use of GeSi NCs that show higher thermodynamic stability than Ge NCs [82, 83], and from the use of multilayer floating gate that ensures better control of variable programming and retention by multiple tunneling in the floating gate at the same time with reduced leakage in the uppermost layer by Coulomb blockade and leakage decrease with the increase of layers number $n$ [99].

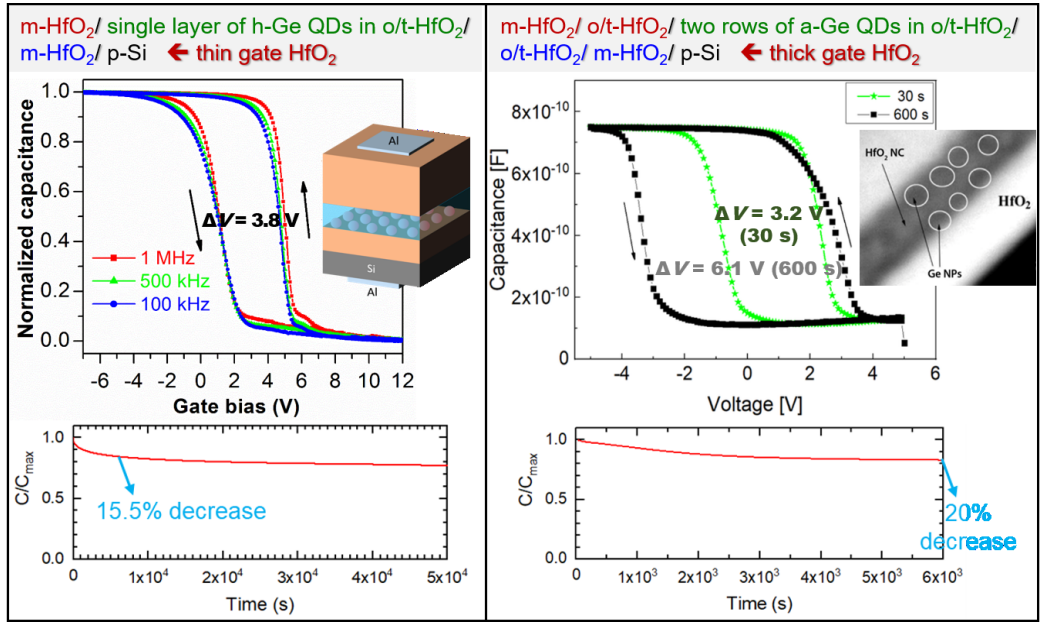

**Fig. 14.** Memory performance ($C – V$ hysteresis loops and charge retention $C – t$ curve) of *HfO$_2$/ Ge-HfO$_2$/ HfO$_2$/ p-Si* NVMs with layers thickness of 22/ 7–8/ 8 nm and 64/ 7–7.5/ 11 nm respectively. m-, h-, o/t- and a- stand for monoclinic, hexagonal, orthorhombic/tetragonal and amorphous structures [96, 97]. © IOP Publishing. Reproduced with permission. All rights reserved.

**Table 3.** Memory performance of 3-layer NVM structures.

| NVM structure (Thickness) / Reference | RTA (°C) | ΔV (V) | Retention | Stored electrons density (cm$^{-2}$) |
|---|---|---|---|---|
| HfO$_2$/Ge-HfO$_2$/HfO$_2$/p-Si (64/7–7.5/11 nm) [97] | 620 | 6.1 | 20% decrease of $C$ after 6000 s | 6.3×10$^{12}$ |
| HfO$_2$/Ge-HfO$_2$/HfO$_2$/p-Si (22/7–8/8 nm) [96] | 600 | 3.8±0.5 | 16% decrease of $C$ after 6000 s<br>14% decrease of $C$ after 3000 – 4000 s<br>≈50% $C$ after 10 years (extrapolation) | N/A |
| HfO$_2$/Ge/HfO$_2$/p-Si (22/5–7/10 nm) [95] | 600 | ≈1 | 28% decrease of $C$ after 4000 s | N/A |
| Al$_2$O$_3$/Ge/[Al$_2$O$_3$/SiO$_2$]/p-Si (10/15/[4/3] nm)<br>(8/11/[4/4.5] nm) [88] | 600<br>650 | 5.4<br>5.1 | N/A | 2.8×10$^{12}$<br>2.6×10$^{12}$ |
| Al$_2$O$_3$/Ge/[Al$_2$O$_3$/SiO$_2$]/p-Si (12/18–20/[4–6/4–6] nm) [88] | 700 | 4.2 | 5% decrease of $V_{FB}$ after 10000 s (ON)<br>4% decrease of $V_{FB}$ after 10000 s (OFF)<br>≈11% loss (ON) after 10 years extrapolation<br>≈10% loss (OFF) after 10 years extrapolation | 2.1×10$^{12}$ |

The structures with continuous Ge layer, i.e. *HfO$_2$/ Ge/ HfO$_2$/ p-Si* (600 °C RTA) show hysteresis loops with frequency-independent $\Delta V \approx 1$ V, but retention properties are not so good (the capacitance decreases with 28% after 4000 s in the charge retention curve, $C - t$) [95]. The memory window is produced only by Ge NCs acting as charge storage centers (frequency-independent $\Delta V$). Also, by increasing RTA temperature to 850 °C, memory properties are deteriorated ($\Delta V \approx$ 0.2 V). Similar behavior with the increase of RTA temperature, was also evidenced in *HfO$_2$/ Ge-HfO$_2$/ HfO$_2$/ p-Si* NVMs (22 nm gate *HfO$_2$*) structures from 600 to 650 and 700 °C [96].

The NVM operation can be affected by the charging-discharging of traps distributed in HfO$_2$ [97], e.g. *HfO$_2$/ Ge-HfO$_2$/ HfO$_2$/ p-Si* NVMs (64/ 7 – 7.5/ 11 nm) structures. So, in the curves of parallel conductance $G_p$ normalized to the electrode area as a function of bias voltage, $G_p - V$, positions shift of the two conductance peaks with the frequency shows the contribution of traps to the charge storage. The two conduction peaks have positions close to the corresponding values of $V_{FB}$ for each hysteresis branch.

Frequency dispersion of the capacitance in the accumulation regime is usually evidenced for the 3-layer memory capacitors with Ge NCs/NPs/QDs floating gate. The main reason of the frequency dependence of the capacitance is due to series resistance in the memory structure, mostly given by the resistance of the p-Si substrate, as presented within [103]. By simulation of the frequency dispersion of capacitance and resistance and comparison with experimental curves measured in the accumulation regime, intrinsic material parameters corresponding to the component layers (dielectric constants and resistivities) are obtained.

*Al$_2$O$_3$/ Ge/ [Al$_2$O$_3$/SiO$_2$]/ p-Si* NVMs present the maximum memory window for RTA in the 600 – 700 °C temperature range (Figure 15, Table 3) [88]. As shown by HRTEM, this is the temperature range in which the density of Ge NCs and NPs reach the maximum. For higher temperatures the strong Ge diffusion reduces their number, while for lower annealing temperatures the Ge amorphous phase is dominant over the crystalline one. So, structures annealed at 600 and 650 °C (amorphous) show the highest memory windows, i.e. 5.4 and 5.1 V respectively due to the high density of Ge related states as deep energy levels. The stored electron densities are also the highest, i.e. $2.8 \times 10^{12}$ and $2.6 \times 10^{12}$ electrons/cm$^2$ respectively (Table 3). The NVMs annealed at 700 °C (majority of Ge in amorphous mixed oxide Ge-rich NPs at the interface and inside adjacent Al$_2$O$_3$ layers, some Ge NCs in floating gate) present hysteresis loops with $\Delta V_{FB} = 4.2$ V slightly lower due to the reducing of the density of localized storage centers. The stored electron densities remain on the order of $2 \times 10^{12}$ electrons/cm$^2$. The retention characteristics, i.e. $V_{FB} - t$ for ON and OFF states are very good, corresponding to the state of the art for NCs floating gate NVMs. So, the charge loss ratio is 5.1% and 4.2%, respectively after 10000 s, and by extrapolation to 10 years is 11% and 9.8%.

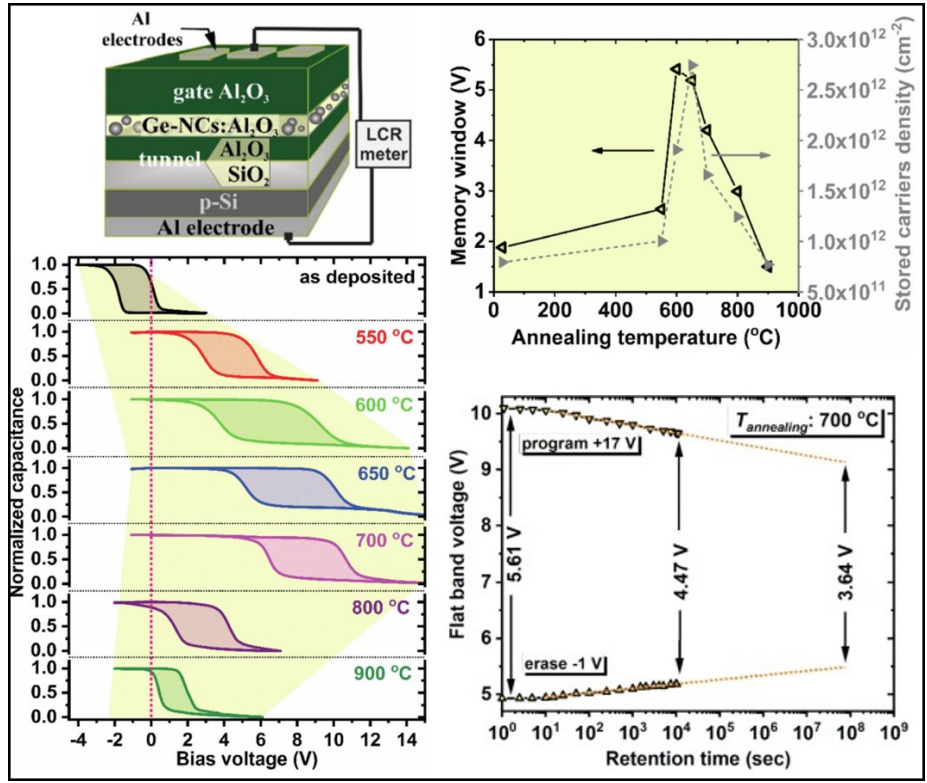

**Fig. 15.** Memory performance of *Al₂O₃/ Ge/ [Al₂O₃/SiO₂]/ p-Si* NVMs: *C – V* hysteresis loops (bottom left) obtained for as-deposited and annealed structures (schematic – top left image) together with results related to memory window and stored electron density (top right); retention characteristic obtained on structures annealed at 700 °C (bottom right) [88]. Reprinted from Appl. Surf. Sci. 542, 148702 (2021), I. Stavarache, O. Cojocaru, V. A. Maraloiu, V. S. Teodorescu, T. Stoica, M. L. Ciurea, Effects of Ge-related storage centers formation in Al₂O₃ enhancing the performance of floating gate memories, Copyright (2021), with permission from Elsevier.

In the case of *SiO₂/ Ge/ SiO₂/ p-Si* structures, NVM devices as cross-bar memories formed of 5568 (100 × 100 μm² cell size) and 871 (300 × 300 μm² cell size) memory cells were fabricated [94]. *C – V* hysteresis loops with memory window *ΔV* up to 2 V were obtained.

Based on the Ge NCs/QDs/NPs floating gate memory structures, different applications were developed, such as photoelectric capacitor memories that combine an optical sensor with an electronic NVM in a single device for photonic flash memories [104], or MOS dosimeters to be used as radiation sensors in space or for medical applications [97]. The benefit from photoelectric NVMs consists in the simultaneous electrical and optical control of the memory effect, thus facilitating the charge tunneling between Ge NCs/QDs/NPs and Si substrate. We reported photoelectric NVMs with photo-induced charging of up to 1.6 electrons/NC and photosensitivity of up to 110 mV/mJ. The NVMs working as dosimeters for the detection of ionizing radiation have high sensitivity (0.8 mV/Gy) to α particle irradiation at low doses (< 50 Gy). Their sensitivity is given by the change in the flatband potential during exposure with α particles in the curve of flatband voltage shift versus absorbed dose after electrical writing of the NVM (charging the floating gate with electrons).

Targeting further improvement of memory properties, besides GeSi NCs in $HfO_2$ multilayered floating gate approach [102], other approach [105] is to use Ge NCs alloyed with both Sn and Si. So, using storage nodes of SiGeSn is advantageous due to the low thermal budget for NCs formation by Sn presence, thus improving the suitability of NVM fabrication in terms of CMOS compatibility, at the same time with ensuring high stability by alloying with little Si. Also, the NC composition can be adjusted, and additionally due to alloying with Sn, highly SWIR photosensitivity of structures can be achieved [85, 86] making possible the fabrication of photoelectric NVMs.

**Concluding remarks**

The great asset of nc-PS is its bright PL in VIS at RT. We have shown, for the first time, that the broad band of PL in VIS contains two subbands only, positioned at 1.54 eV and 1.72 eV, and they are due to the QC effect in NWs and also to surface states. This is specific to Si skeleton with nanometric diameters remained after anodization process. Depending on skeleton morphology and size, properties of nc-PS can differ very much, and therefore they can be tuned by properly adjusting the preparation parameters, and therefore films with desired properties can be obtained.

Ge NCs/NPs embedded in oxides ($SiO_2$, $TiO_2$) show high photosensitivity in VIS-NIR-SWIR at RT in the spectral photocurrent up to 1325 nm for Ge NPs-$SiO_2$ films. Photosensitive Ge NPs-$SiO_2$ films are only achieved by heating the Si substrate during MS deposition. The one-step fabrication by films deposition on heated Si substrate at relatively low temperature withought post deposition RTA is highly advantageous. The sandwich *ITO/Ge NPs-$SiO_2$/n-Si/Al* structures show very good photosensitivity, having high responsivity of 2.42 A/W and 100%-higher internal quantum efficiency of 445%,

*Ge NCs-$TiO_2$/$SiO_2$/n-Si* structures show 600 – 1250 nm spectral sensitivity. The films were obtained by MS followed by RTA (550 ºC). The photocurrent increases exponentially with the bias voltage due to the electrostatic doping by the field effect.

NVMs with floating gate of Ge NCs/NPs/QDs in oxides (classical $SiO_2$ and high high-κ $HfO_2$ and $Al_2O_3$) present high memory performance, the retention characteristics for *$Al_2O_3$/ Ge/ [$Al_2O_3$/$SiO_2$]/ p-Si* NVMs (700 ºC RTA) corresponding to the state of the art for NCs floating gate NVMs, i.e. charge loss of about 10% by extrapolation to 10 years. The memory performance (e.g. memory window $\Delta V_{FB}$) was enhanced by different approaches: i) by preparing a floating gate of a single layer of unstacked Ge QDs with precise separation distance ($\Delta V_{FB}$ = 3.8 V); ii) by increasing the density of Ge-related storage centers as Ge-related local electronic states in addition to Ge-rich NCs ($\Delta V_{FB}$ = 5.4 V); iii) by preparing a floating gate of bilayer stack of Ge NP and iv) by cumulatively adding the contribution of ferroelectric orthorhombic $HfO_2$ to the memory effect ($\Delta V_{FB}$ = 6.1 V and highest stored density of $6.3 \times 10^{12}$ electrons/cm$^2$).

**Acknowledgment**

The research was funded by CNCS – UEFISCDI, projects no. PN-III-P4-ID-PCE-2020-1673, PN-III-P2-2.1-PED-2019-4468 within PNCDI III and by Romanian Ministry of Research, Innovation and Digitalization, NIMP Core Program PN19-03 Contract no. 21N/2019.

# R E F E R E N C E S

[1] V. Reboud *et al.*, Prog. Cryst. Growth Ch. **63**, 1 (2017).

[2] G. M. Xia, Sci. Bull. **64**, 1436 (2019).

[3] S. Barth *et al.*, Chem. Mater. **32**, 2703 (2020).

[4] Y. Miao *et al.*, Nanomaterials **11**, 2556 (2021).

[5] D. Caimi *et al.*, Jpn. J. Appl. Phys. **60**, SB0801 (2021).

[6] V. Raj *et al.*, J. Phys. D Appl. Phys. **55**, 143002 (2021).

[7] D. Valentovic *et al.*, Phys. Stat. Sol. A **56**, 341 (1979).

[8] A. Lesiak *et al,* Nanomaterials **9**, 192 (2019).

[9] O. Toma *et al.*, Appl. Surf. Sci. **478**, 831 (2019).

[10] O. Toma *et al.*, Nanomaterials **11**, 2841 (2021).

[11] L. Ion *et al.,* Proc. Rom. Acad. – Series A **22**, 25 (2021).

[12] D. Manica *et al.*, Nanomaterials **11**, 2286 (2021).

[13] S. Lin and X. Peng, ACS Energy Fuels **35**, 18928 (2021).

[14] E. T. Efaz *et al.*, Eng. Res. Express **3**, 032001 (2021).

[15]   U. e Kalsoom *et al.*, Front. Phys. **9**, 612070 (2021).

[16]   M. Barbato *et al.*, J. Phys. D: Appl. Phys. **54**, 333002 (2021).

[17]   M.-Y. Su *et al.*, Rare Met. (2022).

[18]   B. B. Tesfamariam and J. Ramulu, Mater. Res. Express **7**, 066202 (2020).

[19]   Q. Zhuang *et al.*, Nanoscale Adv. **3**, 3643 (2021).

[20]   X. Li *et al.*, Nanophotonics **10**, 2001 (2020).

[21]   C. Ballif *et al.*, Nat. Rev. Mat. (2022).

[22]   S. Shi *et al.*, Appl. Phys. Lett. **117**, 251105 (2020).

[23]   V. Dhyani *et al.*, IEEE Trans. Electron Devices **67**, 558 (2020).

[24]   J. W. John *et al.*, Nanotechnology **32**, 315205 (2021).

[25]   S. Shi *et al.*, Appl. Phys. Lett. **119**, 221108 (2021).

[26]   Y. Zhao *et al.*, Nanotechnology **31**, 145602 (2020).

[27]   M. Tkalčević *et al.*, Sol. Energ. Mat. Sol. C. **218**, 110722 (2020).

[28]   L. Basioli *et al.*, ACS Appl. Nano Mater. **3**, 8640 (2020).

[29]   H. R. Sully *et al.*, ACS Appl. Nano Mater. **3**, 5410 (2020).

[30]   Y. Li *et al.*, Nanotechnology **31**, 385603 (2020).

[31]   N. Zhang *et al.*, Nanoscale **12**, 13137 (2020).

[32]   Q. Chen *et al.*, ACS Appl. Nano Mater. **4**, 897 (2021).

[33]   M. Mederos *et al.*, Thin Solid Films **611**, 39 (2016).

[34]   D. Lehninger *et al.*, Phys. Stat. Sol. A **215**, 1701028 (2018).

[35]   C. Bonafos *et al.*, *Nano-composite MOx materials for NVMs* in *Metal Oxides for Non-volatile Memory* (Elsevier, Amsterdam, The Netherlands, 2022) pp. 201–244.

[36]   V. Iancu and M. L. Ciurea, Solid State Electron. **42**, 1893 (1998).

[37]   Ed. D. J. Lockwood, *Light Emission in Silicon: From Physics to Devices – Semiconductors and Semimetals* Vol. 49 (Academic Press, San Diego, USA, 2011).

[38]   F. Priolo *et al.*, Nat. Nanotechnol. **9**, 19 (2014).

[39]   Ed. L. Canham, *Handbook of Porous Silicon, 2$^{nd}$ edition* (Springer International Publishing AG, Cham, Switzerland, 2018) part I, pp. 3-99; part II pp. 283-309.

[40]   E. M. L. D. de Jong *et al.*, Light. Sci. Appl. **7**, 17133 (2018).

[41]   N. X. Chung *et al.*, ACS Photonics **5**, 2843 (2018).

[42]   Z. Ni *et al.*, Mat. Sci. Eng. R. **138**, 85 (2019).

[43]   V. V. Nikolaev *et al.*, J. Phys. Chem. C **123**, 27854 (2019).

[44]   L. T. Canham, Appl. Phys. Lett. **57**, 1046 (1990).

[45]  V. Lehmann and V. Gösele, Appl. Phys. Lett. **58**, 856 (1991).
[46]  M. A. Cardona-Castro *et al.*, Nanotechnology **27**, 235502 (2016).
[47]  J.-C. Lin *et al.*, Opt. Mater. Express **7**, 880 (2017).
[48]  S. Chakrabarty *et al.*, IEEE Trans. Device Mater. Reliab. **18**, 620 (2018).
[49]  K. S. Chan and T. J. E. Dwight, Results Phys. **10**, 5 (2018).
[50]  F. Morales-Morales *et al.*, Opt. Mater. **90**, 220 (2019).
[51]  T. Sarkar *et al.*, Appl. Phys. A **128**, 336 (2022).
[52]  D. A. LaVan *et al.*, Nat. Biotechnol. **21**, 1184 (2003).
[53]  T. Sarkar *et al.*, Mater. Res. Express **6**, 115078 (2019).
[54]  M. L. Ciurea *et al.*, Phys. Stat. Sol. B **195**, 637 (1996).
[55]  M. L. Ciurea *et al.*, J. Electrochem. Soc. **146**, 3516 (1999).
[56]  V. Iancu *et al.*, J. Optoelectron. Adv. Mater. **6**, 53 (2004).
[57]  M. L. Ciurea, J. Optoelectron. Adv. Mater. **7**, 2341 (2005).
[58]  M. L. Ciurea, J. Optoelectron. Adv. Mater. **8**, 13 (2006).
[59]  V. Iancu *et al.*, J. Optoelectron. Adv. Mater. **8**, 2156 (2006).
[60]  M. L. Ciurea *et al.*, J. Luminesc. **102**, 492 (2003).
[61]  M. L. Ciurea *et al.*, Chem. Phys. Lett. **423**, 225 (2006).
[62]  A. M. Lepadatu *et al.*, J. Appl. Phys. **107**, 033721 (2010).
[63]  V. Iancu *et al.*, J. Nanopart. Res. **13**, 1605 (2011).
[64]  I. Stavarache and M. L. Ciurea, J. Optoelectron. Adv. Mater. **9**, 2644 (2007).
[65]  M. L. Ciurea *et al.*, Appl. Phys. Lett. **76**, 3067 (2000).
[66]  M. Draghici *et al.*, Phys. Stat. Sol. A **182**, 239 (2000).
[67]  V. Ioannou-Sougleridis *et al.*, Mater. Sci. Eng. C **15**, 45 (2001).
[68]  V. Iancu *et al.*, J. Appl. Phys. **94**, 216 (2003).
[69]  M. L. Ciurea *et al.*, Solid-State Electron. **51**, 1328 (2007).
[70]  C. Popa *et al.*, J. Optoelectron. Adv. Mater. **10**, 2319 (2008).
[71]  A.-M. Lepadatu *et al.*, Sci. Rep. **8**, 4898 (2018).
[72]  I. Stavarache *et al.*, Sci. Rep. **9**, 10286 (2019).
[73]  A.-M. Lepadatu *et al.*, J. Nanopart. Res. **15**, 1981 (2013).
[74]  I. Stavarache *et al.*, Appl. Surf. Sci. **309**, 168 (2014).
[75]  A.-M. Lepadatu *et al.*, Dig. J. Nanomater. Bios. **6**, 67 (2011).
[76]  I. Stavarache *et al.*, J. Nanopart. Res. **13**, 221 (2011).

[77]   V. S. Teodorescu *et al.*, Dig. J. Nanomater. Bios. **8**, 1771 (2013).

[78]   E. Gorokhov *et al.*, *Laser Pulses – Theory, Technology, and Applications* (InTech, Rijeka, Croatia, 2012), pp. 383-436.

[79]   M. L. Ciurea and A. M. Lepadatu, Dig. J. Nanomater. Bios. **10**, 59 (2015).

[80]   H. Lüth, *Solid Surfaces, Interfaces and Thin Films* (Springer-Verlag, Berlin Heidelberg, 2010), p. 372.

[81]   X. Guo *et al.*, Optica **3**, 1066 (2016).

[82]   A.-M. Lepadatu *et al.*, J. Phys. Chem. C **124**, 25043 (2020).

[83]   C. Palade *et al.*, Materials **14**, 7040 (2021).

[84]   M. T. Sultan *et al.*, Appl. Surf. Sci. **479**, 403 (2019).

[85]   A. Slav *et al.*, ACS Appl. Nano Mater. **2**, 3626 (2019).

[86]   A. Slav *et al.*, ACS Appl. Mater. Inter. **12**, 56161 (2020).

[87]   O. Cojocaru *et al.*, Sci. Rep. **11**, 13582 (2021).

[88]   I. Stavarache *et al.*, Appl. Surf. Sci. **542**, 148702 (2021).

[89]   O. Auciello *et al.*, MRS Bulletin **45**, 231 (2020).

[90]   J. Robertson, Eur. Phys. J. Appl. Phys. **28**, 265 (2004).

[91]   K. Mistry *et al.*, *A 45nm Logic Technology with High-k+Metal Gate Transistors, Strained Silicon, 9 Cu Interconnect Layers, 193nm Dry Patterning, and 100% Pb-free Packaging*, 2007 IEEE International Electron Devices Meeting, pp. 247-250 (2007).

[92]   B. Park *et al.*, Microelectron. Eng. **84**, 1627 (2007).

[93]   R. Lo Nigro *et al.*, Materials **15**, 830 (2022).

[94]   D. Vasilache *et al.*, Phys. Stat. Sol. A **213**, 255 (2016).

[95]   A. Slav *et al.*, Scripta Mater. **113**, 135 (2016).

[96]   A. M. Lepadatu *et al.*, Nanotechnology **28**, 175707 (2017).

[97]   C. Palade *et al.*, Nanotechnology **30**, 445501 (2019).

[98]   V. Turchanikov *et al.*, Comparative Studies of Single- and Double-nanocrystal Layer NVM Structures: Charge Accumulation and Retention, Proc. 27[th] International Conference on Microelectronics, pp. 103-104 (2010).

[99]   R. Bar *et al.*, Appl. Phys. Lett. **107**, 093102 (2015).

[100]  S. Shibayama *et al.*, Appl. Phys. Lett. **103**, 082114 (2013).

[101]  M. Dragoman *et al.*, Nanotechnology **31**, 495207 (2020).

[102] Project PCE 191/2021 web page's address https://infim.ro/en/project/multilayered-floating-gate-nonvolatile-memory-device-with-gesi-nanocrystals-nodes-in-nanocrystallized-high-k-hfo2-for-high-efficiency-data-storage-multigesincmem/

[103] C. Palade *et al.*, Appl. Surf. Sci. **428**, 698 (2018).

[104] C. Palade *et al.*, Appl. Phys. Lett. **113**, 213106 (2018).

[105] C. Palade *et al.*, Coatings **12**, 348 (2022).